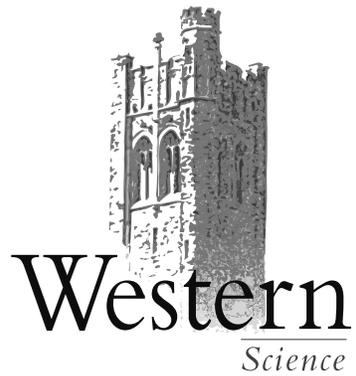

*The* University *of* Western Ontario

# Dynamic Resource Management
# using Operating System-Level Virtualization

*Author:*

Alexander POKLUDA

apoklud@uwo.ca

*Supervisor:*

Dr. Hanan LUTFIYYA

hanan@csd.uwo.ca

March 30, 2010

# Abstract

This thesis expands upon an existing system called Golondrina that performs autonomic workload management among a cluster of hardware nodes running operating system-level virtualization. Golondrina works by identifying localized resource stress situations and attempting to dissipate them by reallocating system resources and, if necessary, migrating or replicating virtual machines. It is predicted that, using Golondrina, efficiency of similar systems can be further improved by achieving greater resource utilization on the hardware nodes while maintaining resource availability for each virtual machine.

The following topics are discussed: virtualization technologies and associated challenges relating to resource management, the architecture and design of Golondrina, intelligent resource reallocation based on predefined policies, and preliminary results demonstrating the effects of a memory resource management policy on the performance of a web application hosted in a virtual environment.

This research makes a significant contribution to the study of virtualized data centres since currently no other system considers virtual machine replication and dynamic memory reallocation as an approach to workload management.

# Acknowledgements

I would like to thank my supervisor, Dr. Hanan Lutfiyya for her support and guidance throughout the project. I would also like to thank Ph.D. candidate Gastón Keller for his mentorship.

# Contents

# List of Figures

# List of Tables

# List of Listings

# Glossary

barrier
: One of the configurable variables for each *user beancounter* parameter. The meaning of barrier differs for each parameter, but it can generally be thought of as a soft limit–that is, a limit that may be exceeded only under special circumstances.

**Beancounter**
: See *User Beancounter*.

**Client**
: The Golondrina run-time component responsible for collecting resource usage statistics on each hardware node and sending these statistics to the *server*. The client is also responsible for adjusting OpenVZ resource allocation parameters and initiating *migrations* and *replications* at the request of the server.

**Container**
: An OpenVZ *virtual environment*. A container in OpenVZ behaves like a stand-alone computer system. Each container can have its own filesystem, memory, hostname(s), IP address(es), system libraries, configuration files, etc. Containers can also have their own super user account and can be rebooted independently. Containers are created using *operating system level-virtualization*.

**Container Image**
: A gzipped tar archive containing the filesystem for a container.

| | |
|---|---|
| `failcnt` | Each *user beancounter* has a corresponding fail count attribute. When an application in a virtual environment requests or attempts to use more of a resource and the use of that resource cannot be granted (either because the container would exceed its limit for that resource or because the resource is unavailable), the value of `failcnt` attribute is increased by one. |
| **Hardware node** | A physical computer running *virtualization* software. |
| `held` | Each *user beancounter* has a corresponding `held` attribute. The value of this attribute is the number of units of a resource that a container is currently using. |
| `limit` | One of the configurable variables for each *user beancounter* parameter. The meaning of `limit` differs for each parameter, but it can generally be thought of as a hard limit–that is, a limit that may never be exceeded under any circumstances. |
| `maxheld` | Each *user beancounter* has a corresponding `maxheld` attribute. The value of this attribute is the maximum number of units of a resource that a container has used at any instant in time since the start of the accounting period (typically since the instance of container was started). |
| `oomguarpages` | An OpenVZ *user beancounter* that is used to guarantee that a certain amount of virtual memory pages are available for applications in a container to use. |
| **OpenVZ** | A technology that implements container-based virtualization using the Linux kernel and GNU/Linux operating system. |
| **Operating System-Level Virtualization** | A form of *platform virtualization* implemented at the operating system level. Operating system-level virtualization enables many isolated virtual environments to be created on a single machine. Since the abstraction of computer resources is performed at the operating system level, all virtual environments share a single kernel. This improves performance and flexibility but requires virtual environments to run the same operating system. For example, OpenVZ virtual environments must run a distribution of the GNU/Linux operating system. |

**Platform Virtualization**  The creation and management of *virtual machines*. Various forms of platform virtualization include full virtualization, paravirtualization, hardware-assisted virtualization and *operating system-level virtualization*.

`privvmpages`  An OpenVZ *user beancounter* that is used to limit the amount of virtual memory pages that applications in a container may allocate.

**Resource Stress**  A situation in which a container or hardware node is using all, or nearly all, of one or more resources available to it, possibly causing degraded performance.

**Server**  The Golondrina run-time component responsible for receiving resource usage statistics from the *clients*. The server is also responsible for analysing these statistics to identify *resource stress* situations. When resource stress situations are identified, the server attempts to dissipate them by instructing clients to adjust resource allocation parameters and/or carry out *migrations* or *replications*.

**User Beancounters**    A set of OpenVZ resource parameters that establish limits and guarantees for various system resources. These parameters are primarily concerned with various aspects of memory usage and are enforced by the kernel. Each parameter has configurable `barrier` and `limit` values.

**Virtual Environment**    A generic term for a *virtual machine.* Different virtualization platforms often use different terms to refer to virtual environments. For example, Xen and VMware use the term virtual machine, while Solaris uses the term *zones* [1], BSD uses the term *jails* [2], Linux-VServer uses the term *virtual private server (VPS)* [3] and OpenVZ uses the term *container* [4]. The term virtual environment is frequently used when discussing operating system-level virtualization.

**Virtual Machine**    A software implementation of a computer that executes programs manner similar to a physical computer.

**Virtualization**    An abstraction of computer resources. Various resources can be abstracted, leading to different forms of virtualization such as application virtualization, network virtualization and platform virtualization. The virtualization technology used in this report, OpenVZ, implements *operating system-level virtualization*, a form of *platform virtualization.*

`vmguarpages`                   An OpenVZ *user beancounter* that is used to
                                guarantee that a certain amount of virtual
                                memory pages may be allocated by applica-
                                tions a container.

# 1   Introduction

The abstraction of computer resources, known as *virtualization*, is a technique that was first used in the 1960s to allow multiple programs to run on a single computer. After the advent of hardware circuits that facilitated the transfer of control between the operating system and one or more independent application programs, research on virtualization slowed. The recent proliferation of inexpensive shared-memory multiprocessor computers along with trends in system administration, however, have lead to a renewed interest in virtualization, and along with it, a new set of challenges has emerged.

*Platform virtualization* enables the execution of one or more *virtual machines* on a single computer. A "guest" operating system runs in each virtual machine and is isolated from every other virtual machine sharing the same resources. Modern methods of platform virtualization can be divided into four main categories: *full virtualization*, *hardware-assisted virtualization*, *paravirtualization* and *operating system-level virtualization*. Full virtualization provides a complete simulation of a hardware architecture. This means any binary code capable of running on the physical hardware can be run within a fully-virtualized environment. Hardware-assisted virtualization provides the same functionality as full virtualization but utilizes specialized hardware features to improve virtual machine performance. Paravirtualization does not simulate hardware so guest operating systems must be modified to use an additional virtualization application programming interface (API) provided by the paravirtualization layer. Paravirtualization generally out performs both full and hardware-assisted virtualization. In operating system-level virtualization, all virtual machines (also known as *virtual environments*) share one operating system kernel. Since neither hardware simulation nor a virtualization API is necessary, operating system-level virtualization has very little overhead and applications executed in a virtual environment can achieve near native performance.

In modern data centres, numerous hardware systems are very often under-utilized and system administrators are turning more and more towards platform virtualization for server consolidation. By using virtual machines to run many independent software systems on a single physical server, greater levels of resource utilization can be achieved [5]. Higher utilization levels mean that fewer physical resources are required to run the data centre and overall costs are reduced [6]. While early research on virtualization focused on sharing the resources of one host computer, recent research has focused on sharing the resources of a cluster of host computers, or *hardware nodes*, among a large number of virtual machines.

## 2    Problem Statement

One of the main challenges in this new prevailing view of virtualization is how to effectively manage the resources of a cluster of hardware nodes as a single unit. Several systems exist or have been proposed that perform autonomic workload management among a cluster of hardware nodes. Such systems typically strive to meet predetermined levels of resource availability for each virtual machine, commonly known as Service Level Objectives (SLOs), while striving to meet predetermined levels of resource utilization on each host. By converging towards these two goals simultaneously, the overall efficiency of a data centre will be maximized while the associated costs are reduced.

This thesis expands upon an existing system called Golondrina. Golondrina is a system that performs autonomic workload management among a cluster of hardware nodes running operating system-level virtualization software. Operating system-level virtualization was selected as the virtualization technology for Golondrina since it provides many fine-grained controls for resource management that are not available in other technologies. Golondrina works by identifying localized *resource stress* situations then attempting to dissipate them by reallocating system resources and, if necessary, by migrating or replicating virtual machines.

The focus of this thesis is a new system based on the original system discussed in [7] that has been architected and implemented by the author. The contributions of this thesis are: memory management using OpenVZ was studied, a simple heuristic for identifying memory resource stress situations was developed, and a new flexible system architecture capable of managing multiple resources simultaneously was created. Moreover, the system was implemented using the Python programming language and the Twisted event-driven networking engine, the original processor resource management code was packaged as a plug-in[1] for the new architecture, and a new memory management plug-in was produced that is capable of performing dynamic resource reallocation based on predefined policies. Documentation for the system was also produced so that other researchers can expand on it. Finally, the system was validated by studying the effects of a memory resource management policy on the performance of a web application hosted in a virtual machine.

It is predicted that, using Golondrina, the efficiency of similar systems can be improved upon by achieving greater resource utilization on the hardware nodes while maintaining resource availability for each virtual machine. This research makes a significant contribution to the study of virtualized data centres since currently no other system considers virtual machine replication and dynamic memory reallocation as an approach to workload management.

The systems most similar to this research are 1000 Islands [8] and the framework proposed by [9]. While these systems have objectives and an overall design similar to Golondrina, they differ from this work since neither employs replication nor dynamic memory reallocation. The Transcendent Memory project [10] and VMware ESX Server [11] demonstrate two approaches to distribute a pool of free memory on a hardware node to guests based on a time-varying basis; however, both are limited by the guest operating system's expectation that it will always have a fixed amount of memory available. Since this research is concerned with operating-system level virtualization, another approach is taken and this limitation is avoided.

# 3   OpenVZ

The specific virtualization technology chosen for this project is *OpenVZ*. OpenVZ is a mature open-source community project sponsored by the Swiss company Parallels Incorporated. OpenVZ forms the foundation for Virtuozzo Containers, a proprietary software product that features easy-to-use control panels and management tools in addition to the basic OpenVZ functionality. Both OpenVZ and Virtuozzo containers are widely used in industry; OpenVZ is particularly popular with web hosting companies that sell virtual private servers (VPSs) to customers over the internet.

---

[1]A plug-in is a computer program that interacts with a host application, in this case Golondrina, to provide specific, additional functionality.

OpenVZ implements operating system-level virtualization. An OpenVZ virtual environment is known as a *container*. Operating system-level virtualization was chosen for Golondrina because, as mentioned previously, it provides fine grained controls over resource management that are not present in other forms of virtualization. OpenVZ provides four primary controls for per-container resource accounting and limiting to ensure the resource usage by one container does not impact other containers. These include: 1) *user beancounters*, which are a per-container set of resource guarantees and limits (discussed further below); 2) disk quota management; 3) a CPU fair scheduler; and 4) configurable input/output priorities [12]. System administrators can also use standard Linux resource management and accounting mechanisms, such as the `tc` traffic control utility, to manage per container network bandwidth limits and `iptables` for traffic accounting.

## 3.1   Memory Allocation

Operating system-level virtualization makes it trivial to add or remove arbitrary amounts of memory to or from a container in real time. An analogy on a physical computer would be adding or removing RAM while the system is running. Other forms of virtualization typically have a kernel that is identical, or nearly identical, to the kernel that would run on physical hardware running in each virtual machine, and that kernel expects the amount of physical memory to remain constant.

In the current version of OpenVZ, virtual memory is managed by the kernel at the hardware node level. To the applications in a container it appears as though the system has a certain amount of physical memory and no swap space, while in reality, some of the container's virtual memory pages may have been swapped out to disk by the kernel. Future versions of OpenVZ may allow a system administrator to set a guaranteed minimum amount of physical memory for a container that shall not be swapped to disk.

In OpenVZ, the mechanism used to add or remove memory to or from a container is the user beancounters.

### 3.1.1    User Beancounters

User beancounters are the primary mechanism of resource management in OpenVZ. There are currently 24 such parameters and each parameter has five attributes: `held`, `maxheld`, `barrier`, `limit` and `failcnt`. These values can be set using the vzctl command on the hardware node and read from the virtual file `/proc/user_beancounters`. On the hardware node, this file lists the user beancounter values for all running containers and the hardware node itself, while within a container it lists the beancounter values for that container only. The units of each parameter depend on the resource being accounted; for parameters relating directly to memory, the primary unit of measurement is pages (4 KiB on a 32-bit machine). The `held` value is the number of units of a resource currently being used by a container, the `maxheld` value is the maximum number of units of a resource a container has used at any one time since the start of the last accounting period (typically since the container instance was started). The `barrier` is typically a soft limit; in general low-priority resource allocations above this limit may be denied in the event of an overall resource shortage on the hardware node while high-priority resource allocations have a greater chance of succeeding up to the `limit` value, which is a hard limit: all resource allocation attempts above this limit within a container are guaranteed to fail. The `failcnt` is an integer field that counts the number of times a resource allocation attempt was denied by the kernel.

Tables 1-3 describe the three main user beancounters that relate directly to memory management: `vmguarpages` ("guaranteed virtual memory pages"), `privvmpages` ("private virtual memory pages"[2]) and `oomguarpages` ("out-of-memory guaranteed pages"). The `barrier` value of the `vmguarpages` beancounter is used to set a memory allocation guarantee–all memory allocation attempts by applications in a container

---

[2]A private memory page can only be accessed by the application that owns it (and the kernel).

**`vmguarpages` User Beancounter**

| Variable | Description |
| --- | --- |
| `held` | not used |
| `maxheld` | not used |
| `barrier` | memory allocation guarantee; all memory allocation attempts below this limit are guaranteed to succeed |
| `limit` | not used |
| `failcnt` | not used |

Table 1: The different variables associated with the **`vmguarpages`** user beancounter. This is the the primary parameter used to allocate system memory to containers.

are guaranteed to succeed if the total memory allocated by those applications is currently below this value. The `privvmpages` beancounter is used to set soft and hard limits for memory allocation. The `oomguarpages` beancounter is used to establish a memory guarantee and is used only in the event of an out-of-memory situation on the hardware node. If the amount of virtual memory used by all of a container's processes is below this value when an out-of-memory situation occurs on the hardware node, then the container is guaranteed that *none* of its processes will be killed; otherwise, the container's processes are at risk of being killed by the kernel out-of-memory killer. The left side of Figure 1 shows the typical relative settings of the these parameters. The User Beancounter section of the OpenVZ wiki [13] and the OpenVZ User's Guide [14] provide more detailed information about the user beancounters.

Other user beancounters which relate to memory are described below.

- `physpages`: The `held` variable of the `physpages` user beancounter indicates the current number of virtual memory pages used by processes in a container that are currently stored in physical memory. (This value does not include any pages belonging to processes in the container that have been swapped out to disk by the kernel at the hardware node level). The `barrier` and `limit` variables are currently unused. In future versions of OpenVZ, the `barrier` and `limit` may be used to limit the amount of physical memory that processes in

### `privvmpages` User Beancounter

| Variable | Description |
| --- | --- |
| `held` | the current number of private (or potentially private) virtual memory pages allocated (but not necessarily used) |
| `maxheld` | the maximum number of private (or potentially private) virtual memory pages allocated at one time since the start of the current accounting period |
| `barrier` | soft-limit for memory allocation; normal priority allocations (e.g. malloc calls by user processes) below this limit will success if the hardware node has sufficient available resources while above this limit they will always fail but higher priority allocations (e.g. process stack expansion) have a higher chance of succeeding |
| `limit` | hard-limit for memory allocation; all allocation attempts above this range will fail |
| `failcnt` | the number of failed allocation attempts since the start of the current accounting period |

Table 2: The different variables associated with the `privvmpages` user beancounter. This is is an auxiliary parameter used to allocate system memory to containers.

<div align="center">

**`oomguarpages` User Beancounter**
</div>

| Variable | Description |
|---|---|
| `held` | the current number of virtual memory pages actually used (includes RAM and swap) |
| `maxheld` | the maximum number of virtual memory pages used at one time since the start of the current accounting period |
| `barrier` | memory guarantee; if the total number of virtual memory pages actually used by all the container's processes (`oomguarpages held`) plus used kernel memory and buffers is below this limit when an out-of-memory situation occurs on the hardware node, then this container is guaranteed that *none* of its processes will be killed, otherwise the container's processes are at risk of being killed by the kernel out-of-memory killer; the OpenVZ User's Guide states this value should be calculated from the `vmguarpages barrier` (above) |
| `limit` | not used |
| `failcnt` | this value is incremented whenever this container has a process killed by the kernel out-of-memory killer |

Table 3: The different variables associated with the `oomguarpages` parameter. This is a secondary parameter used to allocate system memory to containers.

a container may use at any one time.

- `kmemsize`: The `held` variable of the `kmemsize` user beancounter indicates the amount of kernel memory used by the container, in bytes. The `barrier` and `limit` variables are used to limit the amount of kernel memory that may be used by the container.

- `tcpsndbuf`, `tcprcvbuf`, `othersockbuf`, `dgramrcvbuf`: These user beancounters can be used to determine and limit the amount of buffer memory used by each container.

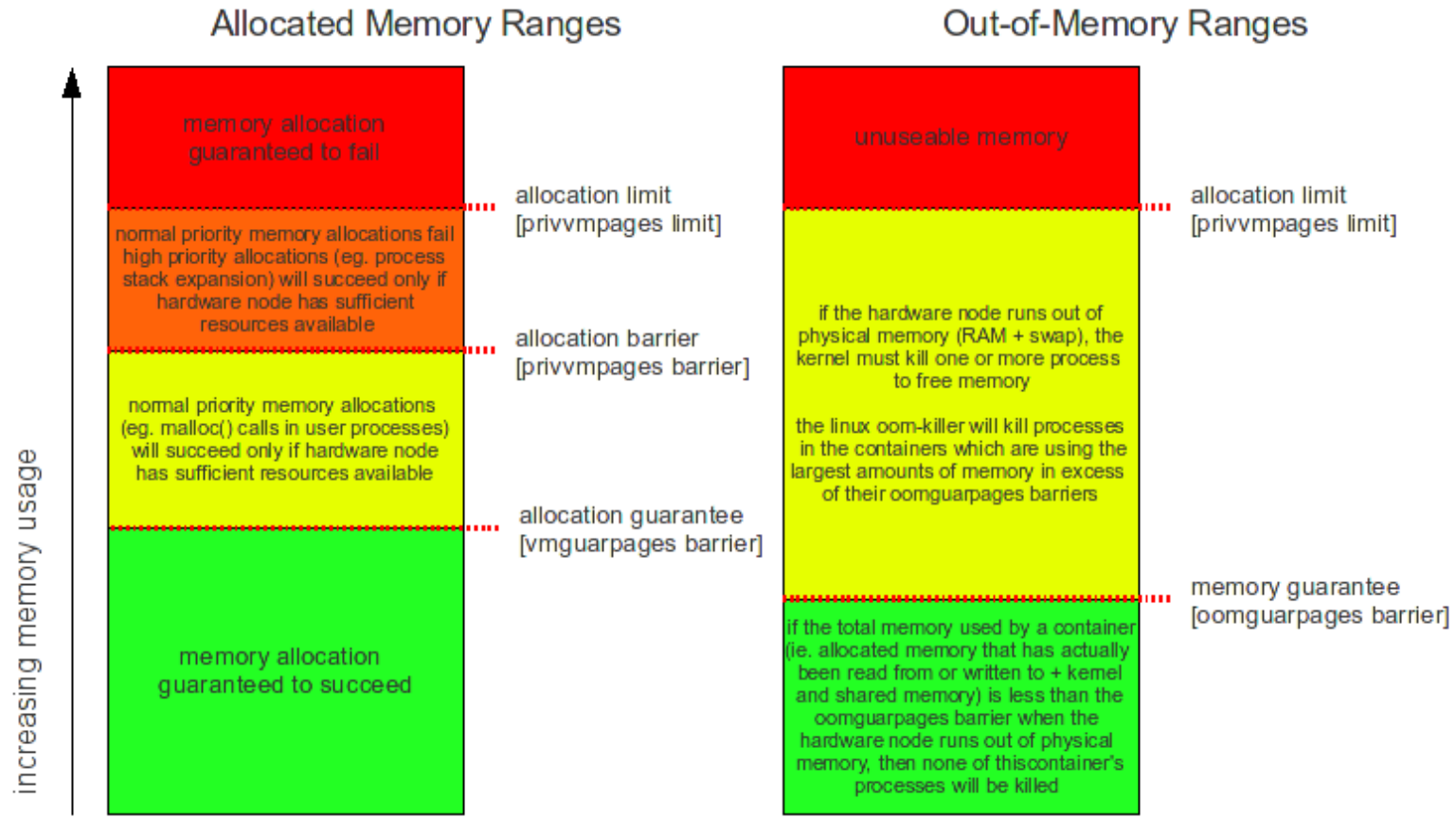

Figure 1: Typical relative settings of the `privvmpages`, `vmguarpages` and `oomguarpages` user beancounters for a container. The left side of the figure describes how the user beancounters are used under normal operation and the right side of the figure shows how the user beancounters are used when an out-of-memory situation occurs on the hardware node.

### 3.1.2   Out-of-Memory Situations

Linux and OpenVZ use demand paging (physical memory is not actually allocated until it is used) and allow virtual memory overcommitment by default. When virtual memory is overcommitted, the sum of all memory allocation guarantees or limits for the containers on a hardware node exceeds the total physical memory and swap space on the hardware node. It may be desirable to have some overcommitment since programs rarely use all the memory they allocate and it is unlikely that all of the processes on a hardware node will try to use the memory guaranteed to them at the same time [15]. Thus, without overcommitment, it may be hard to achieve high levels of memory and swap utilization. When a process first tries to access allocated memory, the kernel must find some physical memory that is not currently in use, or free memory by moving some data from memory to disk. If the kernel discovers that no physical memory is or can be made available, an out-of-memory situation occurs and the kernel may attempt to recover memory to keep the system operational by killing one or more processes.

OpenVZ is distributed with tools that can be used by administrators to help ensure that the resources guaranteed to containers on a hardware node do not excessively exceed the physical resources available by verifying that all the user beancounter barriers and limits are set in a consistent and stable configuration. A stable configuration is one in which it is unlikely the kernel will ever need to kill a process to recover memory. Instead, when memory gets scarce, processes will be given an opportunity to handle out-of-memory situations gracefully (for example, through failed `malloc()` calls).

Golondrina takes samples of resource usage statistics at regular intervals and dynamically reallocates resources to keep system stable. This means that resources can be overcommitted, achieving a high level of resource utilization, with less risk of an out-of-memory situation occurring.

Nevertheless, too much overcommitment may result in undesirable behaviour. If the kernel needs to kill one or more processes to recover memory to keep the system operational, the "out-of-memory killer," contained in `mm/oom_kill.c` of the Linux source code, selects a process to kill based on a "badness" score. The process with the highest badness score will be killed first. The badness score is computed based on a point system using the criteria described below [16].

1. The size of all virtual memory areas allocated to a process, and its children, are computed. More points are given for a larger size.

2. Lower priority processes ("nicer" processes in Linux terminology) get more points.

3. Processes running as the root user are assumed to be important, as are processes that interact directly with hardware. Processes which have been running for a longer period of time are also assumed to be important, since these are likely to be daemon or server processes. These processes have their scores reduced.

4. The kernel swapper and init processes, as well as all other kernel threads are immune, and will not be killed.

This heuristic may not be perfect, but it is designed to be fast and operate with "least surprise." The process with the most points is sent the `SIGTERM` signal, which requests that it release any resource it is using and terminate. OpenVZ introduces the additional criterion that processes belonging to containers that are using the largest amount of memory in excess of their `oomguarpages barrier` will be killed first. The right side of Figure 1 shows how the `oomguarpages` user beancounter is used when an out-of-memory situation occurs on a hardware node.

## 3.2    Golondrina and OpenVZ Resource Management

The first version of Golondrina monitored a single resource–the processor. The processor time used by processes in each container was monitored by reading statistics from the virtual file `/proc/vz/vestat`, provided by OpenVZ. Once a resource stress was identified, Golondrina would attempt to dissipate it through one or more migrations or replications. The original processor resource management code has been packaged by the author and included in the current version of Golondrina as a plug-in.

A memory management plug-in has also been developed. This plug-in monitors the memory usage within each container by monitoring the virtual file `/proc/user_beancounters` and in the memory usage on the hardware node overall by analysing the output from the OpenVZ `vzmemcheck` utility. When a memory resource stress is identified, the memory management plug-in first attempts to resolve the resource stress locally by adjusting the user beancounter barriers and limits. If this is not possible, then it attempts to dissipate the resource stress through one or more migrations or replications.

# 4   Managing Resource Stress Situations

The primary goal of Golondrina is to ensure a predetermined level of resource availability for each virtual environment, while striving to achieve high levels of resource utilization on each hardware node. Golondrina does this by identifying localized resource stress situations then attempting to dissipate them by reallocating system resources and, if necessary, by migrating or replicating virtual environments.

Ideally, when no resource stress situations are occurring, the containers should be distributed so that a high level of resource utilization, say $x$, is achieved on each hardware node. This would ensure that the containers are being run with the minimum number of hardware nodes necessary, reducing overall costs. A pool of additional resources should be maintained to handle resource stress situations. When a resource stress occurs, additional resources should be introduced into the system to alleviate the resource stress and the containers should be redistributed across the hardware nodes to restore this configuration.

Distributing the containers across the hardware nodes to achieve a resource utilization level of $x$ on each is equivalent to the NP-complete subset-sum problem described in [17].

*Proof.* Let $S_0$ be the set of containers to be distributed across several hardware nodes. If we wish to achieve a target resource utilization level of $x$ on each hardware node, then we need to find a subset $s_0 \subseteq S_0$, such that the sum of the resource usage of each container in $s_0$ is exactly $x$. Once this subset has been found, the containers in $s_0$ are assigned to a hardware node and the process repeats for $S_1 = S_0 - s_0$. $\diamond$

Finding good resource allocation strategies is challenging and an active area of research. This version of Golondrina has been designed to be a framework for experimentation with different resource allocation strategies to develop heuristics or approximation algorithms. In Golondrina, resource allocation strategies are implemented as policies.

Golondrina provides three primary mechanisms that policy implementations can use to dissipate resource stress situations. First, Golondrina provides an API to adjust the OpenVZ user beancounter barriers and limits. Second, Golondrina provides an API to perform live container migrations. Third, Golondrina provides an API to replicate container instances.

A basic policy implementation for resolving memory resource stress situations has been developed as a part of this thesis. The following subsections describe how this policy detects and resolves memory resource stress situations.

## 4.1   Resource Stress Indicators for Memory

The following are some indications that the memory available to user space processes in a container or on a hardware node is insufficient, in decreasing order of severity, along with the current heuristic for measuring the level of stress employed by the memory management plug-in. Each resource stress indicator is given a raw score, based directly on user beancounter values, and a normalized score between 0 and 1. The overall memory stress score for a container is taken to be the greatest of the individual scores. Roughly speaking, the memory stress score for a container represents the fraction of guaranteed memory resources that it is using.

1.  *An increase in* `oomguarpages failcnt` *value*

    An increase $x$ in the `oomguarpages` fail count value since it was last checked indicates that $x$ processes have been killed due to an overall memory shortage in the hardware node. When an overall memory shortage occurs in the hardware node, the kernel tries to free memory by killing a process in the container which is using the most memory in excess of its guaranteed amount. This is the greatest indicator of a resource stress and every attempt should be made to avoid this situation.

    - **raw score**: the increase in `oomguarpages failcnt`, i.e. the number of processes that have been killed
    - **normalized score** = $\begin{cases} 0.0 & \text{if } rawscore \text{ is } 0 \\ 1.0 & \text{otherwise} \end{cases}$

2.  *An increase in* `privvmpages failcnt` *value*

    An increase $x$ in the `privvmpages` fail count value since it was last checked indicates that $x$ attempts to allocate more memory in that container were denied by the kernel. Clearly, if the container is requesting more memory and the requests are being denied, that container is experiencing a resource stress.

    - **raw score**: the increase in `privvmpages failcnt`, i.e. the number of failed memory allocation attempts
    - **normalized score** = $\begin{cases} 0.0 & \text{if } rawscore \text{ is } 0 \\ 1.0 & \text{otherwise} \end{cases}$

3.  *Current memory usage (*`oomguarpages` held *+* `kmemsize` held *+* *buf* held*) versus* `oomguarpages` barrier

If a container is using more memory than its `oomguarpages barrier` then its processes are at risk of being killed in the event of an overall memory shortage on the hardware node.

- **raw score**: (`oomguarpages held` + `kmemsize held` + `*buf held`) / `oomguarpages barrier`, i.e. the fraction of guaranteed memory used
- **normalized score** = $\min\{1.0, rawscore\}$

4. `privvmpages held` *versus* `privvmpages barrier`

   If the value of `privvmpages held` is greater than the value of `privvmpages barrier`, then this container has allocated more memory than its memory allocation guarantee. Processes in the container are at risk of having future memory allocation attempts denied. Furthermore, it is possible the hardware node will be stressed once processes in the container write to this (excess) allocated memory.

   - **raw score**: `privvmpages held` / `privvmpages barrier`, i.e. the fraction of memory the memory allocation guarantee used
   - **normalized score** = $\min\{1.0, rawscore\}$

The above heuristic has been implemented in the memory memory management plug-in. Memory usage statistics are collected at regular intervals and analysed using the above heuristic to identify memory resource stress situations. A container is considered to be under a memory resource stress when its normalized score exceeds a predefined threshold that is set in the Golondrina global configuration file.

## 4.2   Resource Stress Resolution for Memory

The basic policy implementation for resolving memory resource stress situations that has been developed as a part of this thesis attempts to resolve memory resource stress situations by applying the following heuristic when a container has been identified as experiencing a resource stress: If the hardware node has memory available, the memory limits set by the user beancounter parameters are increased, giving container access to more memory. If that cannot be done, then the stressed container is migrated to another hardware node with sufficient free memory to allow the user beancounter limits to be increased. If this cannot be done either, future versions of the plug-in will attempt to select a different container on the same hardware node and migrate it to another hardware node so that the memory limits of the stressed container can be increased. If the policy has been configured to perform replications, then another instance of the container may be started after the existing container(s) reach a specified level of resource utilization.

When a hardware node is stressed, an attempt is made to migrate the container that is using the largest amount of resources on that hardware node. If this cannot be done without causing a resource stress on another hardware node, then the container with the next largest resource usage is chosen and the process repeats.

Preliminary results from experiments investigating the effects of this policy implementation on the performance of a web application hosted in a virtual environment are presented in Section 10, Experimental Results.

# 5    Architectural Overview

This system has been architected and designed as a research system that will be a test bed for various resource management policies. As such, it has been architected with modifiability and maintainability as its primary quality attributes.

The primary architectural pattern for Golondrina is "client-server" which is described by [18]. An instance of the client component is run on each hardware node that will host virtual machines and one instance of the server component is run on a dedicated hardware node. The client components are responsible for collecting resource usage statistics at regular intervals and sending them to the server over a local area network (LAN). The server component is responsible for receiving the resource usage statistics and analysing them to identify possible resource stress situations. Once a resource stress situation has been identified, the server contacts the client on the same hardware node as the stressed container and instructs it to adjust the resource limits for the container or migrate it to another hardware node. The server may also contact a hardware node and request it to start a new replica of a container. After a replica of a container has been started, another run-time component called the "gate" is responsible for updating a load-balancer's configuration so that incoming requests are evenly distributed between the previously existing container(s) and the new replica.

Figure 2 shows a deployment diagram for Golondrina. The system has been architected to allow new run-time components to be integrated into the system. An administrator graphical user interface (GUI) is shown in the upper left corner of the deployment diagram. This component may be developed and integrated into the system in the future to allow an administrator to monitor the resource allocation in the system and manually reallocate resources.

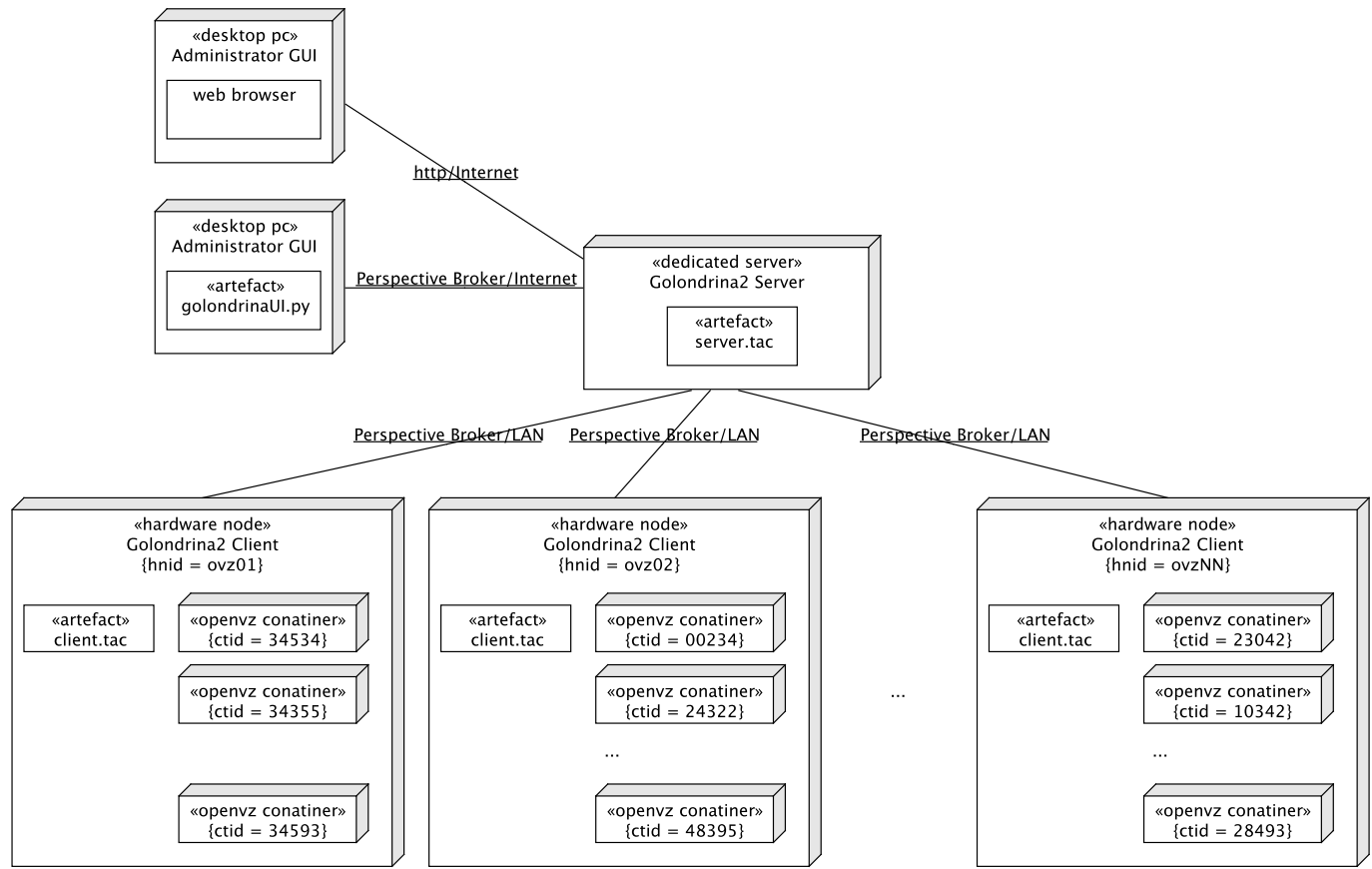

Figure 2: Golondrina deployment diagram. An instance of the client component runs on each hardware node that will host virtual machines to be managed and sends resource usage statistics to a server component. The server component is responsible for managing system-wide resource allocation. The gate (not shown) is responsible for updating a load balancer's configuration after a container replica has been created. (key: UML)

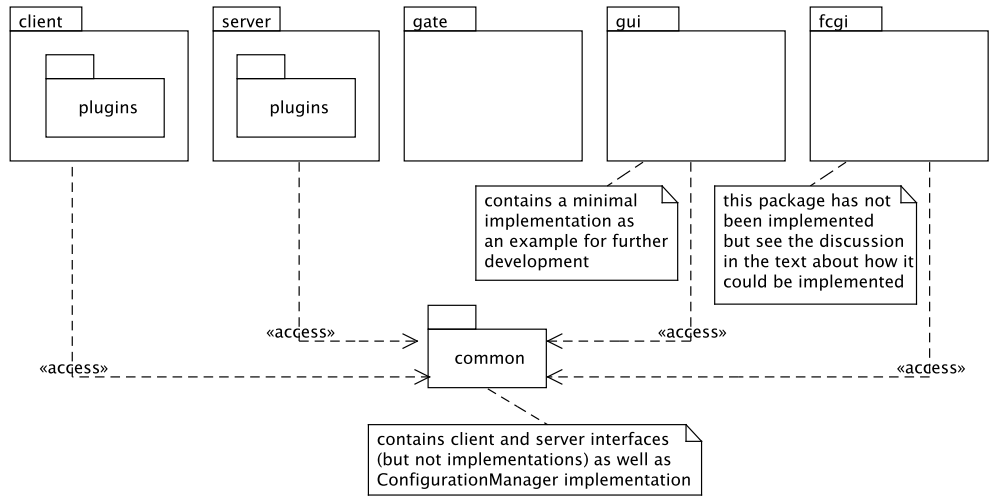

Figure 3: A package diagram for Golondrina. The common package contains code that is used by the other packages. (key: UML)

A package diagram for Golondrina is shown in Figure 3. The code for Golondrina is decomposed into four primary packages: client, server, gate, and common. The common package contains code this is used by the other three primary packages. In order to run the server component, the code from the server and common packages must be present, similarly for the client and gate. One additional package also exists–the GUI package contains a minimal implementation of a standalone desktop application that is capable of communicating with the server component. This has been provided as a basic example of how an administrator GUI could be implemented as a standalone application.

As an alternative to the standalone administrator application, the administrator GUI could be implemented as a web application by developing a fast-common-gateway-interface (FCGI) artefact capable of running within a web server such as the Apache HTTP Server. The fcgi package does not exist in the current version of Golondrina.

The *plug-in* pattern has been used as an architectural sub-pattern and the software responsible for collecting resource usage statistics and analysing them for specific resources is encapsulated in plug-ins. This makes it very easy to modify the system by adding support for managing additional resources. These plug-ins are referred to as resource management or policy plug-ins and their code is placed the plug-ins sub-packages shown in Figure 3.

# 6    Design Overview

The design of Golondrina supports modifiability in many ways. For example, many of the well-known design patterns described by Gamma, Helm, Johnson and Vlissides (commonly known as the "Gang of Four") in [19] have been incorporated into the design. These patterns describe effective solutions to known design challenges and increase the modifiability of the system. Furthermore, the system has been implemented using the Python programming language, enabling changes to the source code to be performed quickly with no recompilation necessary, and the components have been implemented using object oriented methods. The Twisted event driven networking engine has been used to implement the communication between the various run-time components. Twisted is also used to manage the client, server and gate processes.

## 6.1    Client

Figure 4 shows the core classes that make up the client run-time component. The ClientService class uses the *façade pattern* to provide an application programming interface (API) to the sensor and actuator subsystems. The IClientService class acts as an abstraction of the services provided by the ClientService class, enabling the implementation to be easily modified.

The main component of the actuator subsystem, which is responsible for interacting with OpenVZ to adjust resource allocation to containers, is the Actuator class. This class is shown below the ClientService class in Figure 4.

Communication between the various run-time components is accomplished by the use of *avatars*. An avatar is an object that exists on one component, and is controlled remotely by another component. For example, the client component has an avatar object that it controls on the server. The avatar carries out operations on behalf of its remote counterpart. This extra layer enhances the modifiability and security of the system by preventing the direct coupling of the different run-time components and facilitating authentication and authorization. Exactly one instance of the ServerAvatarOnClient class in Figure 4 will exist in each instance of the client component and is remotely controlled by the server. This allows the server to send messages to the client. Each client component has exactly one avatar that exists in the server that it can control using an instance of the ClientAvatarProxy class shown in Figure 5. The relationship between the avatar and avatar proxy objects is shown in Figure 6.

Figure 7 shows the sensor subsystem of the client component. Resource manage-ment plug-ins are decomposed into several smaller plug-ins–one of which is a sensor plug-in. This plug-in runs within the client component and is responsible for gathering usage statistics for a particular resource. A sensor plug-in inherits from an abstract sensor plug-in class or directly from the ISensorPlugin interface. The abstract plug-in class provides default functionality for some of the methods declared in the ISensorPlugin interface. Figure 7 shows both a memory sensor plug-in and a CPU sensor plug-in. The memory sensor plug-in gathers information about the amount of used and available memory for each container running on the hardware node as well as the hardware node overall. The CPU sensor plug-in does the same for the CPU. The SensorManager class, shown at the top of the diagram, is responsible for loading the sensor plug-ins and requesting them to gather statistics at regular intervals. The most important method of the sensor manager is the computeLoad() method which is implemented using the *template method pattern* with the start(), getLoad() and stop() methods of the ISensor interface.

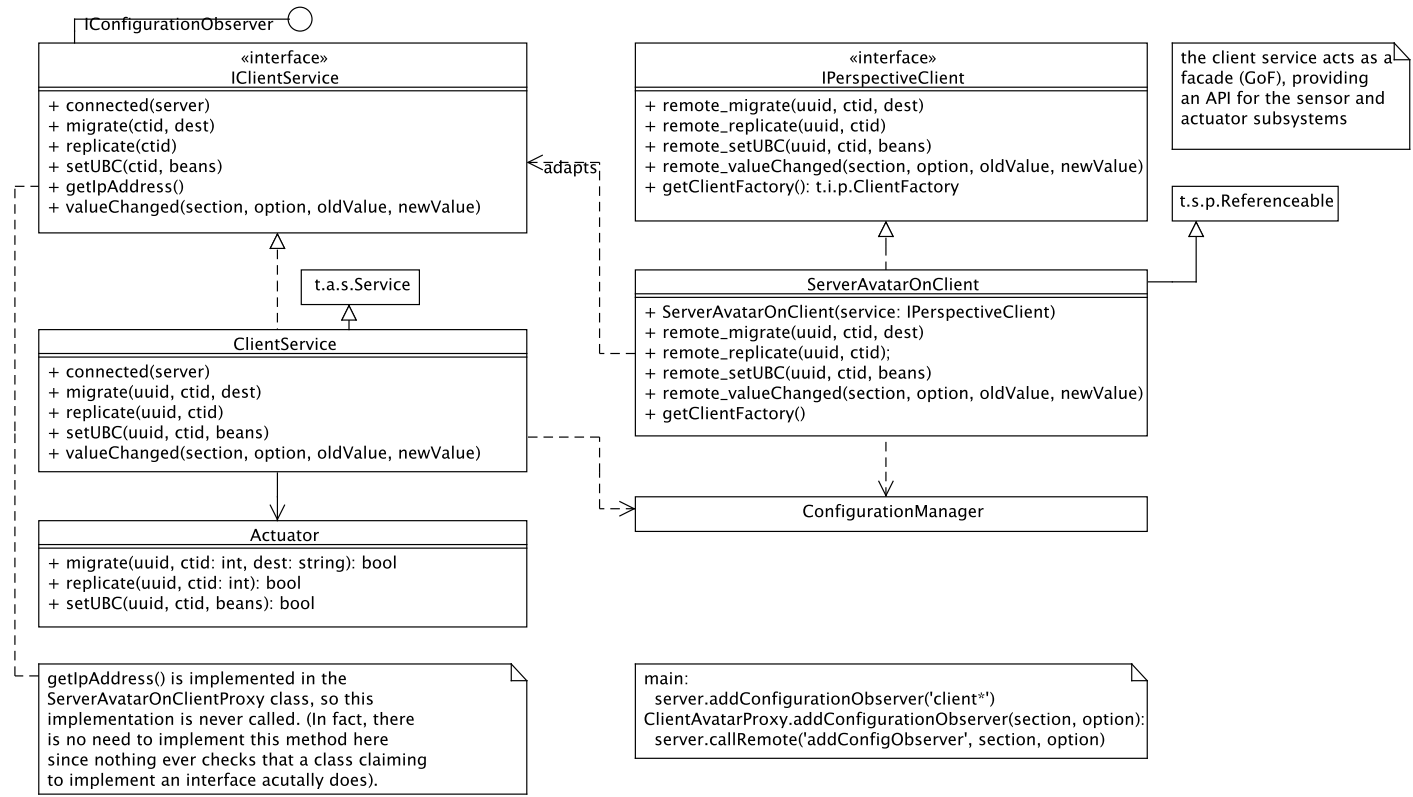

Figure 4: The core classes of the client component. (key: UML)

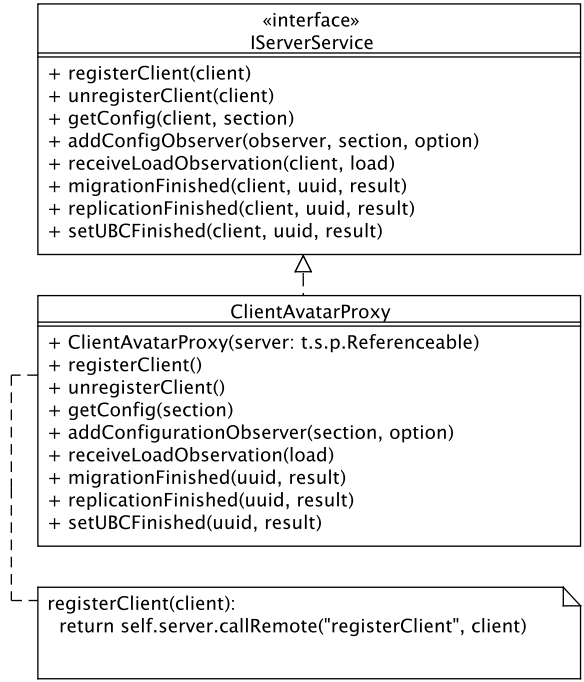

Figure 5: The client avatar proxy class. (key: UML)

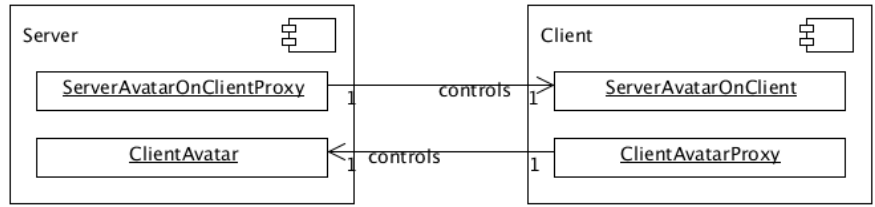

Figure 6: The relationship between the avatar and avatar proxy objects. (key: UML)

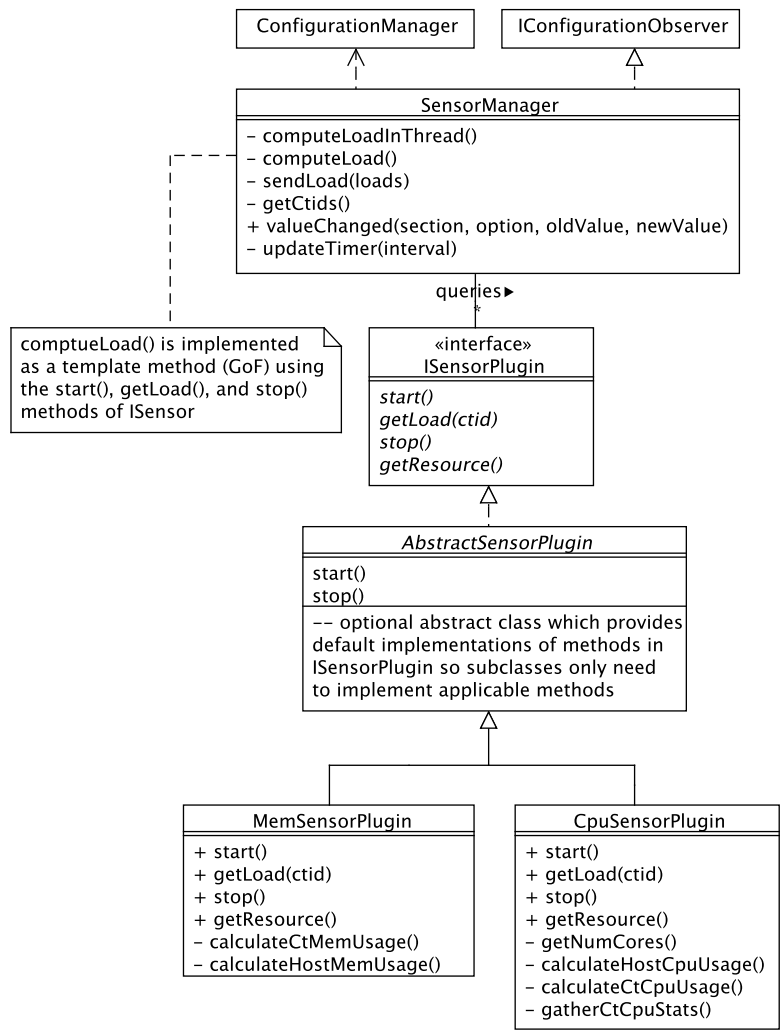

Figure 7: The sensor subsystem of the client component. (key: UML)

All of the Golondrina run-time components have an additional software layer to separate application logic from network code. This layer allows changes to be made to the network protocol that the components use to communicate. Additionally, a new component could be added that transparently uses a different protocol for network communication. Figure 8 demonstrates how the client component could be modified to use an ASCII line-based protocol instead of the remote procedure call protocol that is currently used.

All of the Golondrina run-time components use the configuration manager subsystem to change their run-time behaviour. Figure 9 shows how the client component uses some of the classes that make up this subsystem. The code for this subsystem is located in the common package and its functionality will be discussed further in Section 6.3.

## 6.2 Server

The core classes of the server run-time component are shown in Figure 10. The ServerService class acts as a façade and provides an API for the services that the client component may access.

Usernames and passwords are used by other Golondrina run-time components when they connect to the server to ensure that only authorized components can access the services provided. (The network communication is also encrypted using a secure sockets layer (SSL)). The PerspectiveServerRealm class shown in Figure 10 is responsible for creating avatar objects, such as client avatar objects, for other components connecting to the server after they have been authenticated.

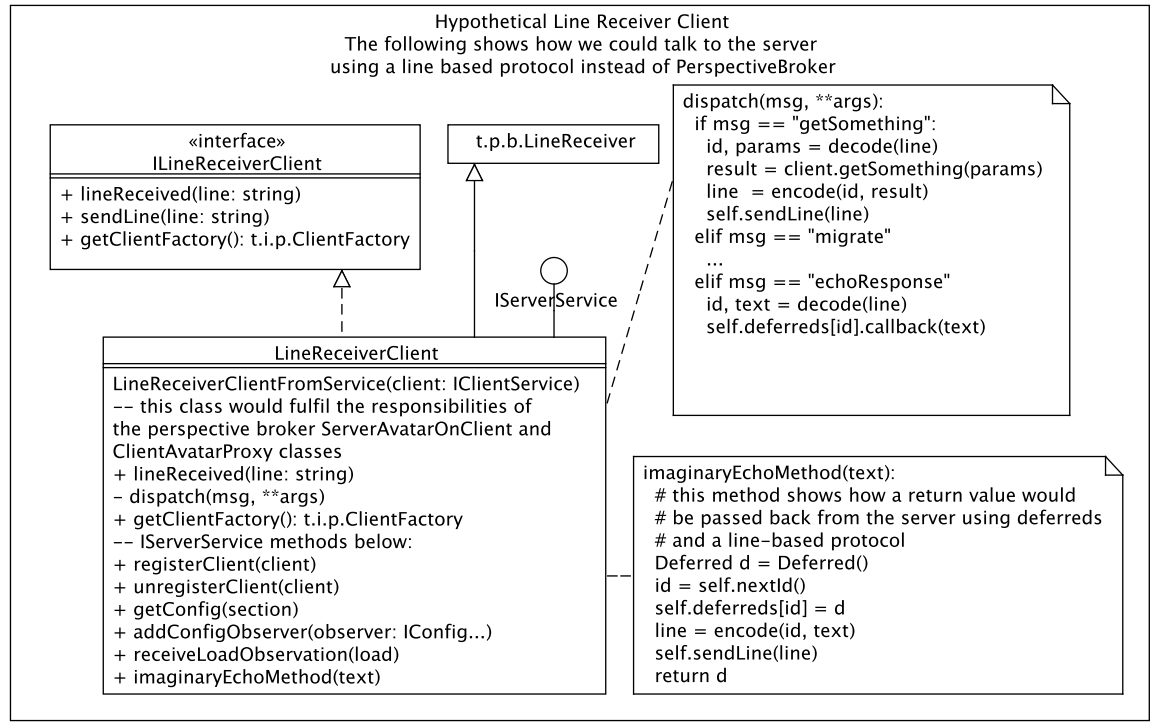

Figure 8: A hypothetical line receiver client. This diagram demonstrates how the client component could be modified to use an ASCII line-based protocol instead of the remote procedure call protocol that is currently used. (key: UML)

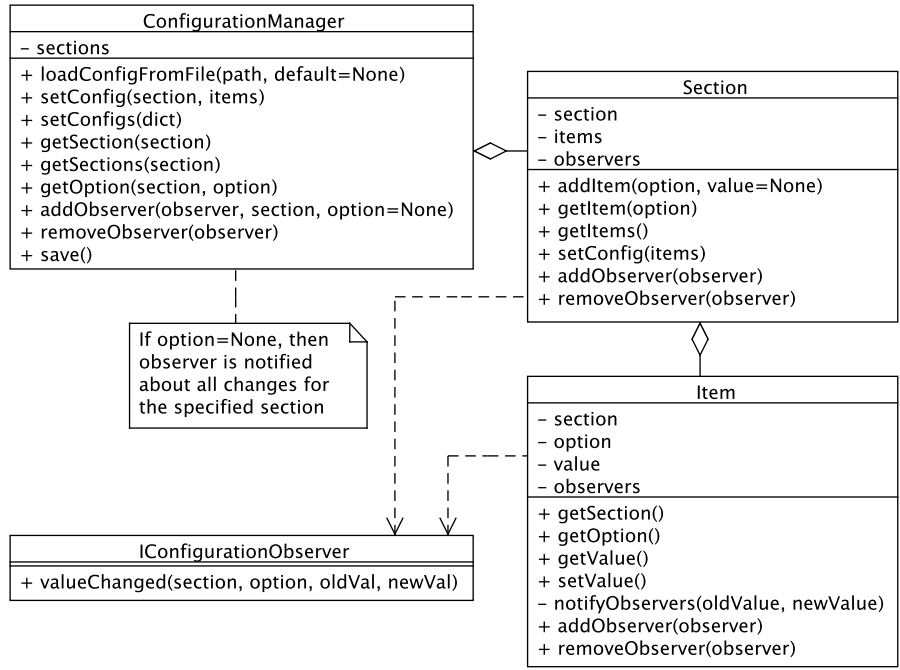

Figure 9: The relationship between the configuration manager classes when they are used by the client. The source code for the configuration manager subsystem is contained in the common package and the same source is used slightly differently by each run-time component. (key: UML)

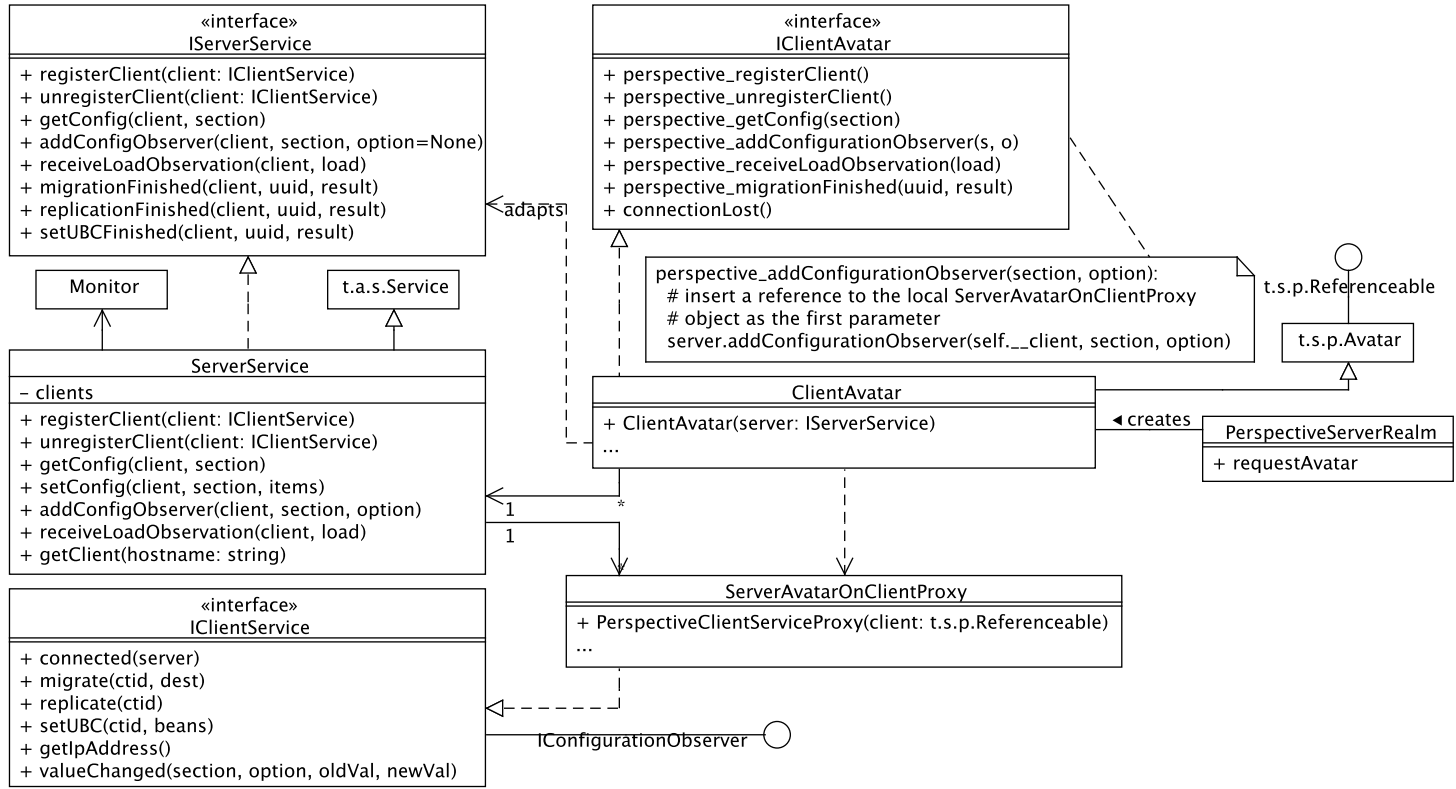

Figure 10: The core classes of the server component. (key: UML)

The ServerAvatarOnClientProxy class is used by the server to send messages to the client components. One instance of the ServerAvatarOnClientProxy class exists for each connected client and they are used to control the ServerAvatarOnClient objects that exist in each client. Likewise, the client uses its ClientAvatarProxy class to control a ClientAvatar object in the server.

As mentioned in the previous section, resource management plug-ins are decomposed into three parts that are loaded independently (these parts can be thought of as plug-ins or "sub-plug-ins" themselves). The first part, the sensor (sub-)plug-in, has already been discussed. The remaining two, which are loaded by the server, are the overload policy plug-in and the overload resolver plug-in. Having resource management plug-ins decomposed into several parts makes modifying a resource management strategy easier. For instance, a memory management plug-in may implement several different policies for determining when a container or hardware node is overloaded. Each of these policies can be encapsulated in different sub-plug-ins, allowing the administrator to mix and match the various algorithms to achieve an optimal resource management strategy.

All of the installed plug-ins will be loaded when Golondrina starts up. Each plug-in is identified by a key such as 'cpu' or 'mem', identifying the resource it manages. Also, since multiple policies may exist for each resource, each sub-plug-in has a unique identifier for that resource such as 'auto_regressive_order_1'. The initial sub-plug-ins to use for each resource may be specified in the global configuration file. The sub-plug-ins may be switched in real time while the system is running by the configuration manager subsystem. The overload policy plug-in system is shown in Figure 11a. The ability to dynamically swap algorithms at run time is known as the *strategy pattern*.

The overload resolver plug-in system has a simpler structure and is shown in Figure 11b. One interesting thing to note is that each overload resolver plug-in is given a priority value. Once a resource stress has been identified, the overload resolver plug-ins for the stressed resource(s) are each given a chance in turn to try and dissipate the resource stress. The overload resolver plug-ins with a higher priority are given an opportunity to resolve the overload before plug-ins with a lower priority. If two plug-ins have the same priority, then the order in which they are run is unspecified.

While policy logic is contained in classes implemented as plug-ins, it may be desirable to change some of a policy's state at run time (a threshold value, for instance). Policy plug-ins have their state encapsulated and abstracted using the *memento pattern*.

The PolicyPluginRepository, which is responsible for managing the policy plug-ins, will call IPolicyStateLoader.getState(resource, id) to get an initial state for each policy plug-in when it is first loaded. The policy state loader is responsible for listening for changes at run time and updating policy states by calling PolicyPlug-inRepository.setPolicyState(). Different implementations of the policy state loader could load policy state information from different sources, such as the filesystem or a database. There is currently only one policy state loader implementation in Golondrina, and it retrieves policy state information from the configuration manager subsystem. The policy plug-in repository and policy state loader are shown in Figure 12.

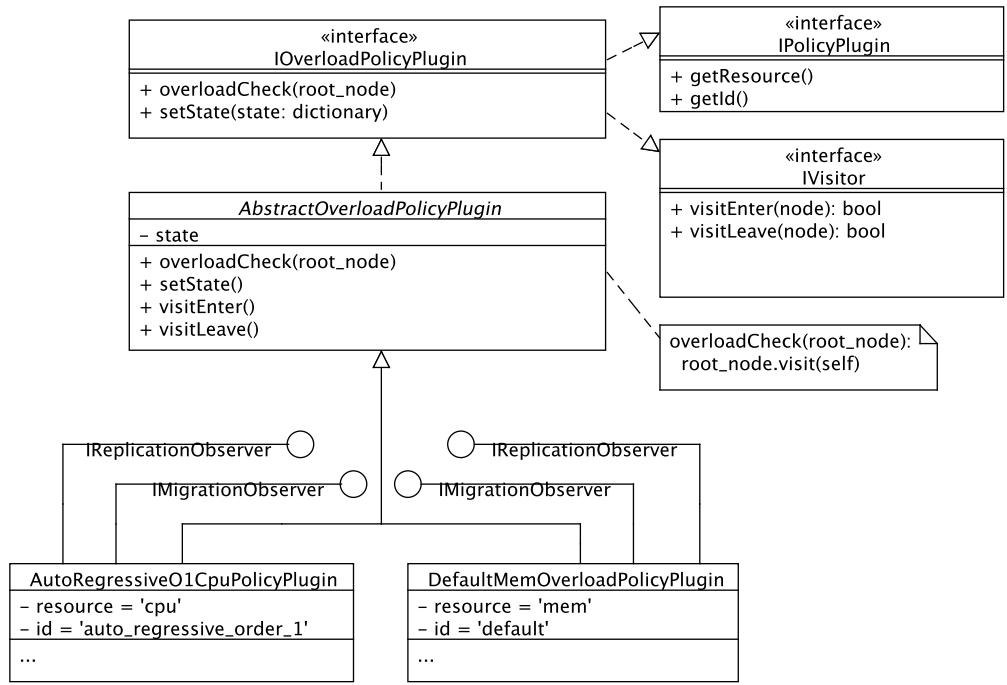

(a) The overload policy plug-in system.

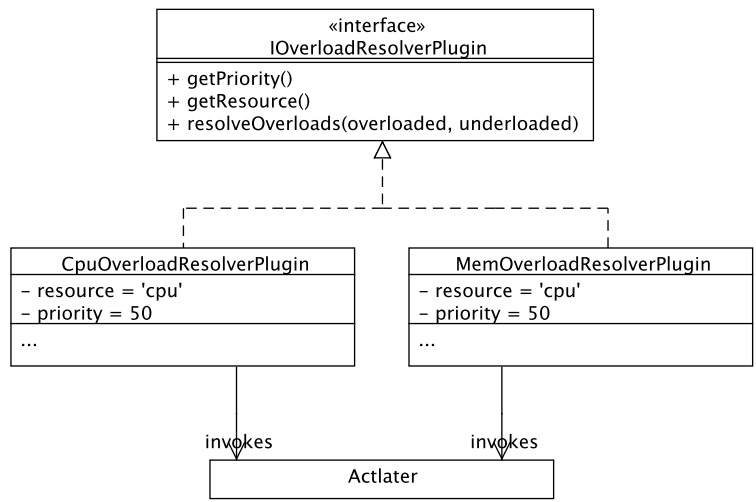

(b) The overload resolver plug-in system.

Figure 11: The overload policy and resolver plug-in systems. Having resource management plug-ins decomposed into several smaller parts makes it easy to modify a resource management strategy. (key: UML)

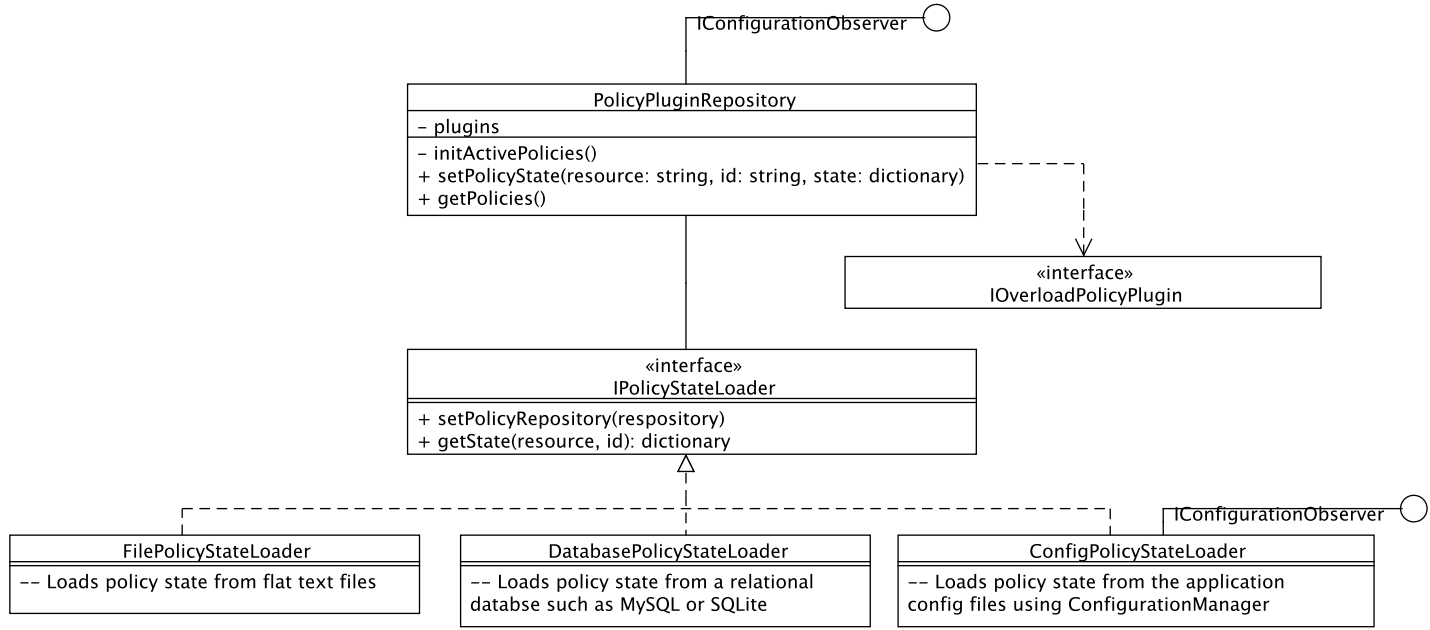

Figure 12: The policy plug-in repository and policy state loader. The policy plug-in repository is responsible for managing the policy plug-ins while the policy state loader is responsible for updating policy states from external data sources. (key: UML)

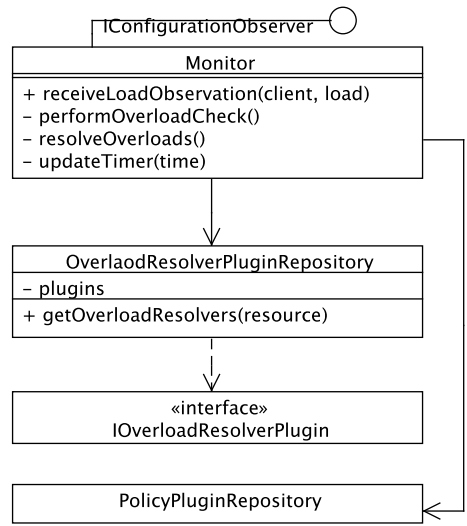

Figure 13: The monitor, which is responsible for receiving load observations from the client components and running periodic resource stress checks using the overload policy plug-ins. If a resource stress is found, the monitor attempts to dissipate it using the overload resolver plug-ins. (key: UML)

Like for the overload policy plug-ins, there is also a repository for the overload resolver plug-ins. This repository is simpler and shown in Figure 13 along with the monitor. The monitor is responsible for receiving load observations (resource usage statistics) from the client components and running periodic resource stress checks using the overload policy plug-ins. The frequency at which to run the stress checks is specified in the global configuration file and may be updated at run time using the configuration manager subsystem. If a resource stress is found, the monitor attempts to dissipate it using the overload resolver plug-ins.

The monitor class shown in the previous figure maintains a data structure that parallels the physical hardware, shown in Figure 14. In this data structure, nodes are used to represent the data centre, hardware nodes and containers. The resource usage statistics that the monitor receives from client components are stored in these nodes. An interface layer is used so that different node hierarchies may be implemented that store load data differently, such as in memory or in a database. Different implementations may also implement different solutions for long-term resource usage statistics storage. The current in-memory node implementation simply discards resource usage statistics after a short period, but it may be desirable to persist these statistics for long term trend monitoring. Operations are performed on this data structure (including data access) using the *hierarchical visitor pattern*. Like the iterator pattern, this pattern has the advantage that visiting objects need not know how the actual data structure is implemented.

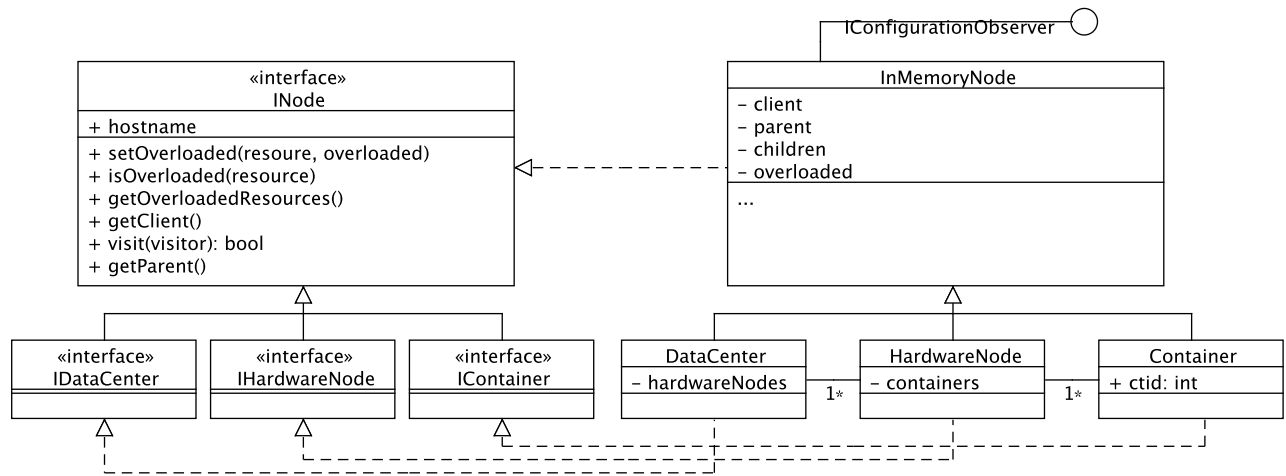

Figure 14: The node data structure hierarchy used by the monitor to store resource usage statistics. (key: UML)

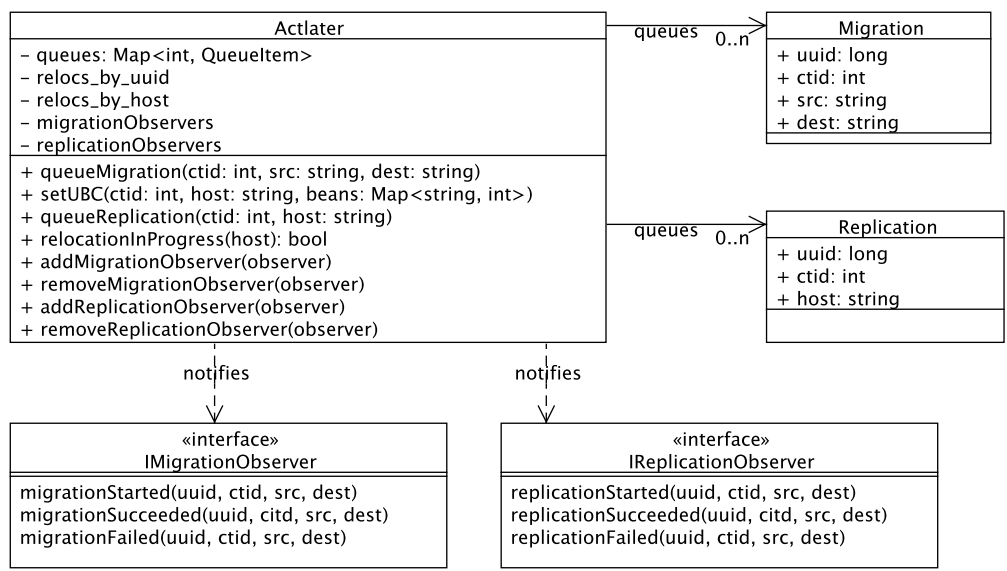

Figure 15: The actlater module is responsible for forwarding instructions to adjust user beancounter parameters to clients and queueing migration and replication operations. (key: UML)

Golondrina provides three primary mechanisms that overload resolver plug-ins may use to dissipate resource stress situations. The overload resolver plug-ins may adjust the user beancounter parameters locally on a hardware node, migrate a container from one hardware node to another, or replicate a container (this is accomplished by instructing a client component to retrieve a *container image* from a repository and start it up). The overload resolver plug-ins are able to perform these oper-ations by using the API provided by the actlater module, shown in Figure 15. The actlater module acts as an intermediary between the overload resolver plug-ins and the client components and queues migration and replication operations. The actlater module executes migration and replication operation requests on a first-in, first-out basis ensuring that each hardware node is involved in at most one migration or replication at a time since performing a migration or replication consumes resources on the hardware node(s) involved. Requests to adjust user beancounter parameters are not queued since executing this operation requires negligible resources on the hardware node.

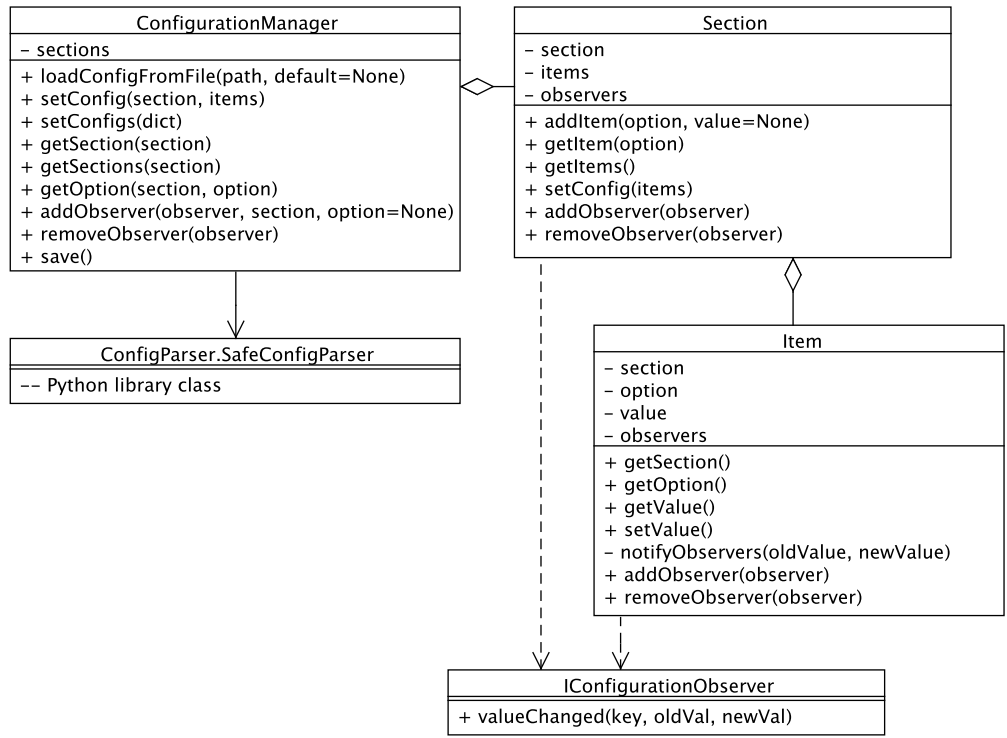

Figure 16: The relationship between the configuration manager classes when they are used by the server. (key: UML)

Users of the actlater module can be notified when a queued operation starts, succeeds or fails by implementing the IMigrationObserver or IReplicationObserver interfaces and adding themselves as a migration or replication observer. This form of notification is an implementation of the *observer pattern*.

The relationship between the configuration manager classes when they are used by the server are shown in Figure 16. On the server, the configuration manager sub-system is initialized by loading a global configuration file using the SafeConfigParser Python library class. The configuration manager subsystem is discussed in the next section.

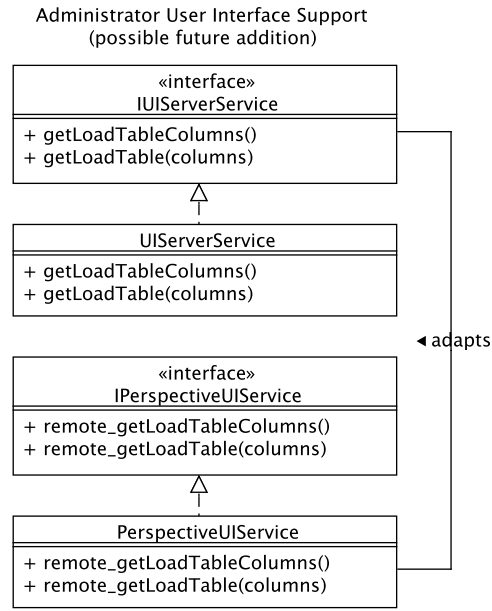

Figure 17: Administrator user interface support (possible future addition).

The ServerService class that was shown in Figure 10 provides an API for the services that the client component may access. If an administrator user interface, such as the one shown in Figure 2, is developed in the future, then a new service class should be created–one that provides an API for services that the administrator user interface may access. An example of such a class is shown in Figure 17.

## 6.3   Configuration Manager

The configuration manager subsystem maintains a database of configuration information for the whole system and can be used to alter the system's run-time behaviour.

The configuration manager stores configuration options in a hierarchy using sections, subsections and values. When the server is started, it first loads default configuration values from the file `server/golondrina-defaults.conf`. It then uses an internal path consisting of the following entries and tries to open each file in turn

1. `server/golondrina-server.conf`

2. `/etc/golondrina-server.conf`

3. `~/.golondrina-server.conf`

Each file that exists and can be read is parsed. Any file later in the path has a higher precedence and any values in that file override values set by previous files. When the client or another component is started and connects to the server, it receives relevant configuration sections. (For example, when a client component connects to the server, it receives a copy of the client configuration section and all of its subsections and options).

Additionally, most configuration values can be updated at run time. When a value is changed at run time, the changes are propagated throughout the entire system, even over the network to other run-time components. (The observer pattern is used again to achieve this effect). Thus, the configuration manager subsystem would make it easy for the administrator user interface mentioned previously to offer an administrator the ability to change parameters that affect the run-time behaviour of the system and have the results of those changes take affect almost instantaneously. Listing 1 shows the current Golondrina global configuration file.

```
1  # GOLONDRINA GLOBAL CONFIGURATION
2
3  [server/general]
4  # port: The port that the server component will use to listen
5  # for connection attempts from other run-time components.
6  # (default: 8888)
7  #port=8888
8
9  [server/ssl]
10 # cert, key: The paths to the SSL server certificate and
11 # private key used for encrypting network communication.
12 cert=certs/ca.uwo.csd.syslab.bravo01.crt
13 key=certs/ca.uwo.csd.syslab.bravo01.key
14
15 [server/data]
16 # max_in_memory_observations: The maximum number of load
17 # observations that the in-memory node class stores in memory
18 # (older observations will be discarded).
19 max_in_memory_observations=20
```

```
20
21  [server/policy]
22  # state_loader: The policy state loader implementation to use.
23  # Possible values which may be implemented are 'file', 'database',
24  # and 'config'.Changes to this value require a system restart to
25  # take effect.
26  state_loader=config
27
28  [server/policy/overload]
29  # check_interval: The number of seconds between overload checks.
30  check_interval=12
31
32  # active_policies: The overload polices to use (selected by
33  # resource and id). You can choose which overload policy to
34  # use dynamically at run time, but if a new plug-in is installed
35  # in the plug-ins folder, it will not be loaded until the
36  # system is restarted.
37  # Note: at most one policy may be selected for each resource.
38  active_policies={'cpu':'auto_regressive_order_1','mem':'default'}
39
40  [server/policy/state]
41  # The initial policy states for the ConfigPolicyStateLoader.
42  # The option and its value will be parsed by the
43  # ConfigPolicyStateLoader. For instance, the line
44  #
45  # overload-cpu-aro1={'threshold':0.75,'foo':'bar'}
46  #
47  # means that the overload policy plug-in with key 'cpu'
48  # and id 'aro1' will be initialized with the state dictionary
49  #
50  #  {'threshold' : 0.75, 'foo' : 'bar'} .
51  #
52  overload-cpu-auto_regressive_order_1={'threshold':0.10}
53  overload-mem-default={'threshold':0.80}
54
55  [client]
56  # frequency: The number of seconds between each round of
```

```
57  # resource usage statistics gathering. (The name 'frequency'
58  # is a misnomer and may be changed in future versions).
59  frequency=2
```

Listing 1: Golondrina global configuration file.

# 7    Implementation Case Study:    Communication Using Avatars

This section investigates the implementation of one aspect of the system: communication using avatars. Figure 18 shows how the client connects to the server and communicates with it shortly after it is started. In this example, we see the client register with the server, receive an initial configuration that is used to populate its copy of the configuration manager, and add itself as a configuration observer on the server so that it is notified of any updates to the configuration values in the client section of the configuration database.

In the upper left hand corner of the diagram, we see the found message getClient-Factory(). When this method is called on a ServerAvatarOnClient object, it creates a ClientAvatarProxy object and a PBReconnectingClientFactory object. The ClientA-vatarProxy object will be used by the client component to control its avatar object on the server once a connection has been established. The ServerAvatarOnClient object then calls __connect() on itself which in turn calls doLogin() on the factory object. The factory object takes care of setting up a connection with the server and providing authentication details. A deferred object is used to notify the ServerA-vatarOnClient object when a connection with the server has been established. The ServerAvatarOnClient object is notified of the successful connection by the deferred object calling its __loginSucceeded() method with a remote reference object that can be used to make remote procedure calls. Next it stores a reference to this object in the ClientAvatarProxy object. The ClientAvatarProxy object can now be used to control a client avatar object on the server that was created when the connection was established.

The ServerAvatarOnClient object then notifies the client service object that a connection has been established. The client service object uses the avatar proxy object to register with the server. (Registering involves server-side initialization). The registerClient() method returns a deferred object immediately, and it is this object that notifies the client service when the registration has completed.

This process is repeated in a similar manner by the client object to retrieve the client section of the configuration database in the server, which is passed to the configuration manager object. The process is repeated one more time with a call to addConfigurationObserver() on the client avatar proxy object. Once this completes, the client service object will be notified about any updates to the client section of the configuration database. Finally, it calls _finalizeInitialization() on itself. This in turn calls __loadSensorManager() which starts the resource usage monitoring.

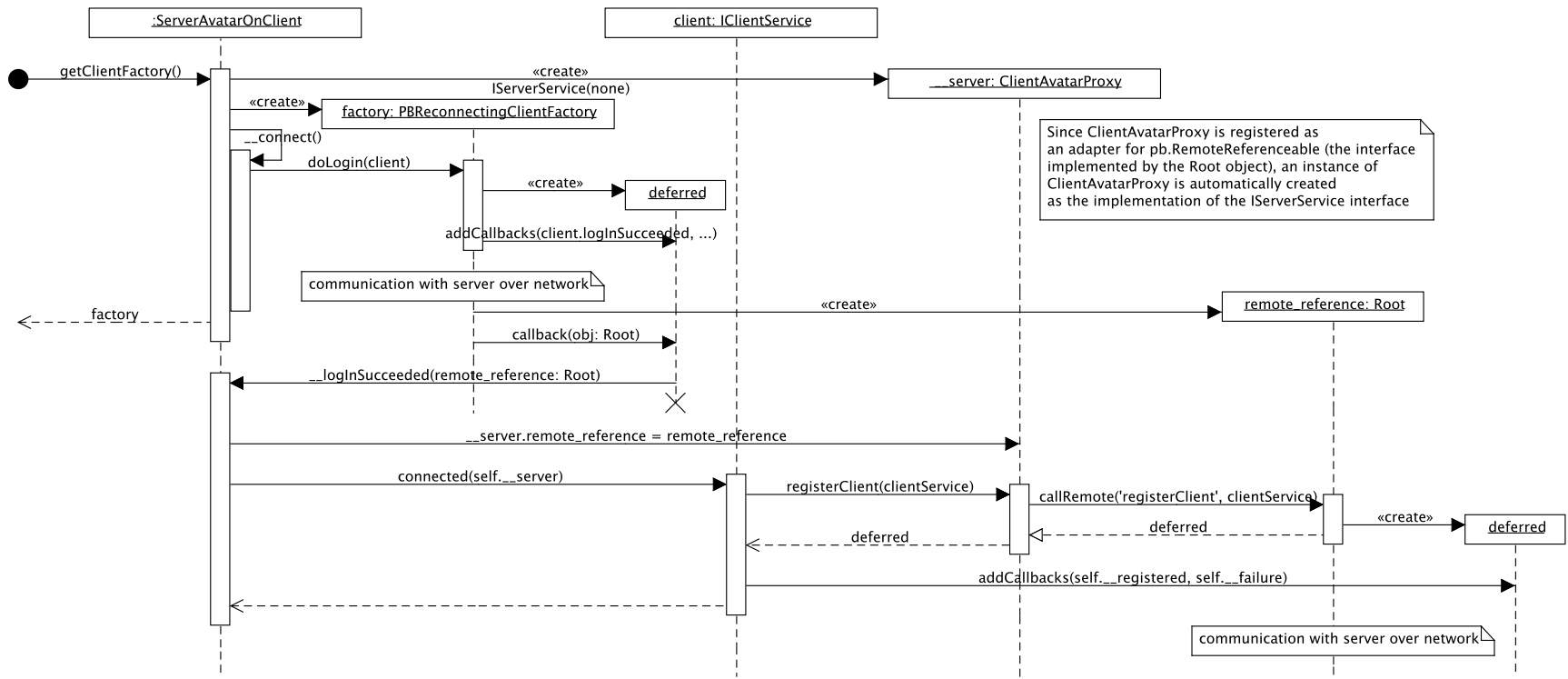

Figure 18: A sequence diagram demonstrating the use of avatars for communication over the network. This specific example shows how the client connects to the server after it is first started. (key: UML)

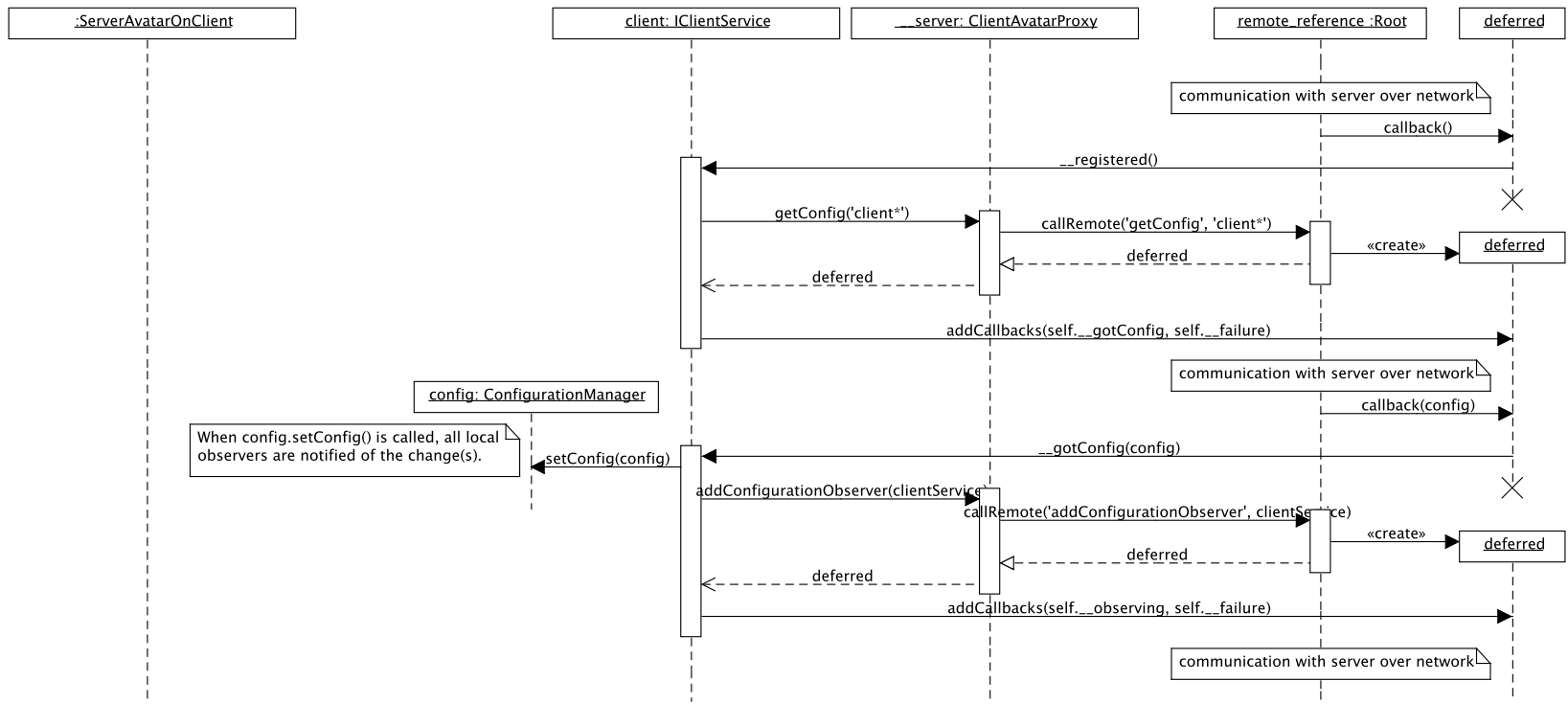

Figure 18: A sequence diagram demonstrating the use of avatars for communication over the network (continued).

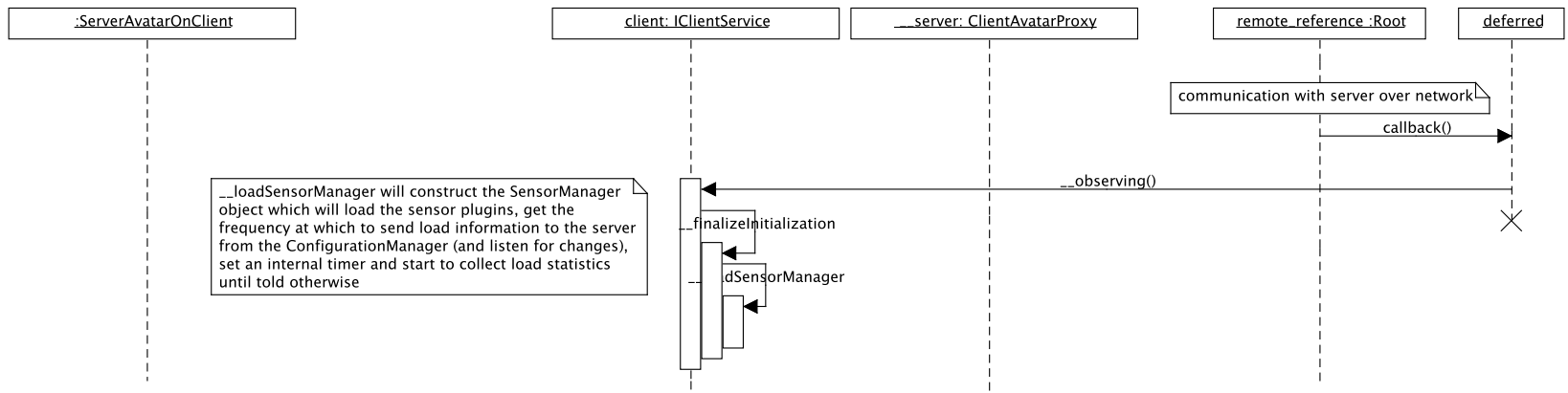

Figure 18: A sequence diagram demonstrating the use of avatars for communication over the network (continued).

# 8   Source Code Availability and Documentation

The source code developed as part of this thesis is free software. Anyone is free to redistribute it and/or modify it under the terms of the GNU General Public Licence as published by the Free Software Foundation, either version 3 of the License, or any later version. A full copy of this license can be obtained from `http://www.gnu.org/licenses/`.

Instructions on how to obtain a copy of the software, as well as instructions on how to set up a development environment, are currently available at `http://alexanderpokluda.ca/trac/cs4490/wiki/Download`.

A significant amount of documentation has been developed so that other researchers can use Golondrina to experiment with various resource management policies. Most of this documentation exists in the form of in-code comments and API documentation. The source code is currently available for browsing online at `http://alexanderpokluda.ca/trac/cs4490/browser/golondrina/trunk`, and the API documentation can be viewed at `http://alexanderpokluda.ca/apidocs/index.html`.

# 9   Test Environment

A test environment was created to demonstrate the effects of using Golondrina for dynamic resource management. Two configurations were investigated for the virtual machines to be stressed (placed under load): a Java implementation of the TPC-W reference benchmark and a web application running on a LAMP solution stack.

The TPC-W benchmark specification is provided by the Transaction Processing Performance Council (TPC). The TPC is a "non-profit corporation founded to define transaction processing and database benchmarks and to disseminate objective, verifiable TPC performance data to the industry" [20]. The TPC-W benchmark describes in detail an e-commerce application and the various ways in which a customer can interact with it. The benchmark also specifies how the performance of an implementation should be measured so that the performance of different implementations running on different hardware can be compared.

A *software* or *solution stack* is a collection of software subsystems needed to deliver a fully-functional product or service. The LAMP solution stack consists of a Linux-based operating system, an Apache HTTP Server, a MySQL database management system and either PHP, Python or Perl interpreters to produce server-side dynamically generated web content. LAMP configurations are used by many of the most popular websites on the Internet, including Wikipedia [21]. The Apache HTTP Server powers over half of the active websites on the Internet according to [22], which is more than three times the number of websites run by the next most popular server.

Joomla! was selected as the specific web application to be tested running on the LAMP stack. Joomla! is a popular, free and open source web content management system written in PHP that provides easy website creation and management. Joomla! provides content template system with modules for creating common website elements such as blogs, forms, polls, news feeds, etc. The Joomla! installer gives the user an option to install a sample website that uses various page layouts as well as many of the default modules, including a news feed, poll and form to demonstrate its functionality. It is this sample website that was the target of HTTP requests generated by a load generator.

Load generation and performance monitoring for the web application running on a LAMP stack was carried out using Apache Jakarta JMeter version 2.3.4. JMeter is a versatile load testing tool designed to work with a variety of web services. JMeter allows the user to create test plans that may include content samplers, delays, and content validation as well as many other items and is capable of producing a variety of reports.

After preliminary experimentation with both configurations, the LAMP configuration was selected for detailed performance analysis because JMeter allows the user to create customized workloads through test plans, while the workloads for the TPC-W benchmark are defined in the specification. It was believed that customizable workloads would facilitate testing different resource management policies.

The containers used in these experiments were created using a CentOS operating system template, based on CentOS version 5.3 (final). After an initial container was created, the Apache HTTP Server version 2.2.3, MySQL database server version 5.0.45 and PHP version 5.1.6 were installed from the CentOS software repositories. Finally, Joomla! version 1.5.14 was downloaded from the Joomla! website [23] and installed with the optional sample website.

Experiments were performed using three identical hardware nodes, bravo01, bravo02 and bravo03, hosted by the Department of Computer Science at the University of Western Ontario. Each hardware node contained a 3.40 GHz dual-core Intel Pentium D CPU with a 2 MiB cache, 2 GiB of RAM and 2 GiB of swap space. bravo01 and bravo03 were running the CentOS enterprise class Linux operating system version 5.3 (final), while bravo02 was running CentOS version 5.2 (final). Each hardware node was running an OpenVZ kernel based on Linux version 2.6.18. bravo01 was running version 2.6.18-128.2.1.el5.028stab064.4 while bravo02 and bravo03 were running

2.6.18-92.1.18.el5.028stab060.2.

After the above mentioned software was installed, as few changes as possible were made the to their default configurations. Unless configuration changes are discussed in the following section, configuration values were only changed where necessary to ensure the interoperability of the different applications.

## 9.1   Apache HTTP Server Configuration

The Apache HTTP Server version 2 contains several *multi-processing modules* (MPMs). The multi-processing modules are responsible for binding to network ports on the server machine, accepting requests and dispatching those requests to children for processing [24]. On Unix and Unix-like platforms such as Linux, the default multi-processing module is the *prefork* module.

The prefork module creates non-threaded HTTP server processes to handle client[3] requests in a manner similar to Apache version 1.3. This module is appropriate for websites that need to use non-thread-safe libraries and offers enhanced security and stability. The prefork module attempts to pre-create, or *pre-fork*, processes before they are needed so that there will always be idle processes ready to handle new client requests as soon as they arrive, avoiding delays in processing associated with process creation [25].

Using this model, there is a one-to-one mapping between client connections and operating system processes. A parent process is responsible for creating child server processes to handle requests, assigning each new client connection to one of those child server process, and recycling child server processes after client connections end.

HTTP/1.1 introduced *persistent connections* to reduce server load and congestion on the Internet [26]. Prior to HTTP/1.1, a separate TCP connection was used for each client request. This would often result in a client establishing several TCP connections to the same web server to download embedded resources in an HTML document. With HTTP/1.1, the same connection may be used for several requests.

The Apache HTTP Server prefork module implements persistent connections by dedicating a server process to each client connection until that connection is closed or until a predefined amount of time has passed. The Apache server has persistent connections enabled by default with a time-out value of five seconds. Persistent connections can be disabled and the time-out value can be configured using the `KeepAlive` and `KeepAliveTimeout` directives in the server configuration files, but the default values were used for the experiments described below.

---

[3]In this section, the term *client* refers to an HTTP client, such as a web browser, not the Golondrina client component.

The number of servers created by the prefork modules can be configured using the `StartServers`, `MinSpareServers`, `MaxSpareServers`, and `MaxClients` directives in the Apache HTTP Server configuration files. The `StartServers` directive defines the number of child server processes to be created after the HTTP server is initially started. The `MinSpareServers` and `MaxSpareServers` directives define the minimum and maximum *idle* child server processes that should exist, ready to handle client requests as soon as they arrive. If the number of idle child server processes is below the value for `MinSpareServers`, the parent server processes creates new child server processes at a maximum rate of one per second until this value is reached. If the number of idle child server processes is above the value for `MaxSpareServers`, then the parent server process kills its children until this value is reached. The `MaxClients` directive sets the maximum number of connections that will be processed simultaneously. The default value for `StartServers` is eight, but it was decreased to four to enable performance testing using fewer resources. Similarly, the default values of `MinSpareServers` and `MaxSpareServers` are five and ten respectively, but these were reduced to two and four. The default value for `MaxClients` is 256 and this value was lowered to 128.

To better understand the behaviour of the prefork module, it is enlightening to consider a simple example. Suppose the Apache HTTP Server has just been started to handle requests for the website at `www.example.com` and that the `StartServers`, `MinSpareServers` and `MaxSpareServers` directives for this server were set using the values given above. Now suppose that Alice opens up her browser and navigates to the `www.example.com` website. When Alice's web browser first connects to the Apache HTTP Server, the parent server process assigns one of the four idle child server processes to handle the connection from Alice's browser. This leaves three idle child server processes to handle future connections, so the parent server process does not create any more.

Suppose that Bob also visits `www.example.com` a couple of seconds after Alice. Since Alice's browser and the server both support persistent connections, the client process assigned to Alice's connection is still in use and is not considered idle, so the parent server process assigns the connection from Bob's browser to one of the other child server processes. Almost immediately after Bob visits `www.example.com`, Carol does too. The parent server assigns the connection to one of the two child processes not in use by Alice or Bob. There is now only one idle client process, so the parent server process creates one more.

Alice, Bob and Carol find the `www.example.com` website to be rather dull and uninteresting, so they each close their web browser and decide to play Solitaire instead of exploring the website further. When they close their browsers, the child server processes that their browsers were connected to become idle again. This brings the number of idle child server processes up to five, so the parent process kills one of them to bring the number of idle processes down to four.

| **Bean** | **Barrier** | **Limit** |
|----------|-------------|-----------|
| `oomguarpages` | memory guarantee | not used |
| `vmguarpages` | memory allocation guarantee | not used |
| `privvmpages` | soft limit for memory allocation | hard limit for memory allocation |

Table 4: A brief summary of the OpenVZ user beancounter parameters adjusted in Tests 1-3.

In the experiments that follow, JMeter threads are used to simulate web browsers belonging to different users.

# 10    Experimental Results

The memory needs of the container configuration described in Section 9 were studied in order to develop a test strategy. It was observed that the default user-space processes in the container, namely init, udevd, syslogd and sshd allocate only about 6 MiB of memory when the container is idle and no users are logged in. The MySQL and Apache HTTP Server processes allocate approximately an additional 30 MiB when they are first started. The top program, which displays a list of tasks currently being managed by the Linux kernel, shows that the virtual memory size of each HTTP server process is approximately 20 MiB, but yet there are five of them (one parent and four children). The reason for this is that much of this memory, such as the memory containing executable code, is shared between server processes. It was also observed that after an Apache HTTP Server child process has served a few requests for PHP pages, the apparent memory allocation of those processes rises to about 30 MiB each. After all the idle child processes have served PHP requests, the memory allocated by the container's user-space processes is approximately 50-60 MiB.

Based on this information, two memory profiles were created: one representing approximately 64 MiB of memory reserved for the container's user-space processes and another representing approximately 128 MiB of memory reserved for the container's user-space processes. These two profiles were created from the "vps.basic" profile provided with OpenVZ (which represents approximately 256 MiB of memory reserved for the container's user-space processes), except the `oomguarpages`, vm-

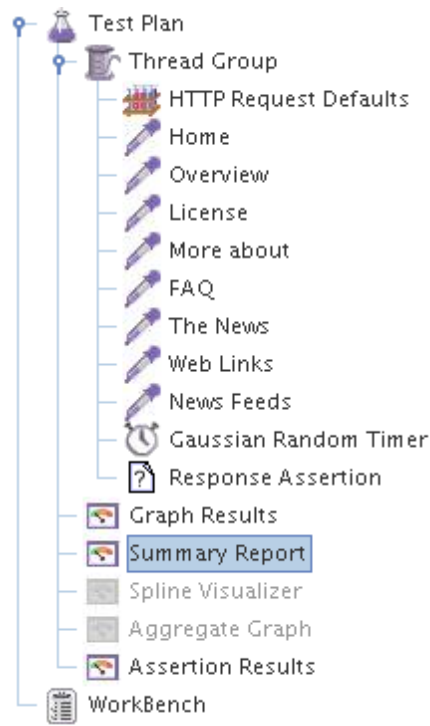

Figure 19: The JMeter test plan outline.

`guarpages` and `privvmpages` user beancounter parameters were set as follows. For the 64 MiB profile, the `oomguarpages barrier` was set to 60 MiB, the `vmguarpages barrier` was set to 64 MiB and the `privvmpages barrier` and `limit` were set to 64 and 66 MiB respectively. For the 128 MiB profile, the `oomguarpages barrier` was set to 100 MiB, the `vmguarpages barrier` was set to 128 MiB and the `privvmpages barrier` and `limit` were set to 128 and 132 MiB respectively. Table 4 provides a brief review of the functionality of these user beancounter parameters.

Figure 19 shows the outline of the JMeter test plan that was developed. The test plan is constructed as a hierarchy of elements with a thread group element and several listener elements at the top level. The thread group is used to control the number of threads JMeter uses to execute a test. The thread group element has several parameters that can be configured, including the number of threads to start, the ramp-up period and a loop count. The ramp-up period, measured in seconds, is used to set the rate at which the test threads are started. For example, if the number of threads is set to ten and the ramp-up period is set to five, then the test threads are started at a rate of two per second for five seconds. The loop count determines the number of times each thread loops.

Below the thread group in the hierarchy is an HTTP Request Defaults element, eight HTTP Request elements, a Gaussian Random Timer element and a Response Assertion element. The HTTP Request Defaults element is used to set default parameters for HTTP Requests at this level and below. In this case, it sets the default hostname for the HTTP Request elements to `ct1891.syslab.csd.uwo.ca`, the fully qualified domain name of the container to which the HTTP requests will be sent. The eight HTTP Request elements are used to request eight different web pages of the Joomla! sample website. For example, the HTTP Request element called Home loads the sample website's home page and the rest are similar. Each iteration of a test thread will load these eight pages in sequence and then repeat until it has been repeated as many times as the loop count value. The Gaussian Random Timer is used to add a short delay at the end of every iteration to reduce the CPU load on the server. The Response Assertion checks that every HTTP response contains a closing HTML tag, `</html>`. This was added as a quick check to ensure that the HTTP response contains a valid HTML web page and not a plain text error message from the PHP interpreter. The Summary Report element provides the average, minimum and maximum response times for each web page, as well as the standard deviation of response times, the number of HTTP responses that did not contain a valid HTML page (as a percentage), the throughput (rate of completed requests), the average bandwidth and the average number of bytes of each HTTP response. The throughput value is computed using the elapsed time between the moment when the first HTTP request is sent and the moment when the last HTTP response is received This includes any delays, such as those introduced by the ramp-up time and Gaussian Random Timer element. Thus, throughput values for tests performed with different ramp-up times cannot be compared directly.

In addition to the parameters discussed above for OpenVZ, the Apache HTTP Server and JMeter, two parameters in the Golondrina configuration file are of interest in the following experiments. These are the threshold value used by the memory management plug-in and the frequency at which the client component collects resource usage statistics. The values for all of the parameters discussed are given in a table at the beginning of each test.

The three hardware nodes, bravo01, bravo02 and bravo03 were used as follows: bravo01 was used to run the Golondrina server component while bravo02 and bravo03 ran the Golondrina client component; bravo01 was also used to run a load generator and performance monitor.

**Initial OpenVZ Settings**

| oomguarpages | 100M:unlimited |
|---|---|
| vmguarpages | 128M:unlimited |
| privvmpages | 128M:132M |

(a) Initial OpenVZ user beancounter barriers and limits for the container that will receive requests from the load generator.

**Apache HTTP Server Settings**

| StartServers | 4 |
|---|---|
| MinSpareServers | 2 |
| MaxSpareServers | 4 |
| MaxClients | 128 |
| KeepAlive | On |
| KeepAliveTimeout | 5 |

(b) Apache HTTP Server settings.

**JMeter Settings**

| Number of Threads (users) | 9 |
|---|---|
| Ramp-Up Period (in seconds) | 2 |
| Loop Count | 5 |

(c) JMeter test plan settings.

**Golondrina Settings**

```
[server/policy/state]
overload-mem-default   {'threshold':0.80}
[client]
frequency                             10
```

(d) Golondrina settings.

Table 5: OpenVZ, Apache HTTP Server, JMeter and Golondrina settings for Test 0. The syntax used for the parameter names and values is the same as that used by each application.

| Run | Avg | Min | Max | Std Dev | Err % | Throughput | Fail Count |
|---|---|---|---|---|---|---|---|
| a | 449 | 167 | 1849 | 251.20 | 0.00 | 8.1 | 0 |
| b | 454 | 166 | 1855 | 235.84 | 0.00 | 7.9 | 0 |
| c | 451 | 166 | 2287 | 241.41 | 0.00 | 8.1 | 0 |
| d | 439 | 166 | 3108 | 246.43 | 0.00 | 7.8 | 0 |
|  | 448 | 166 | 2275 | 243.72 | 0.00 | 8.0 | 0 |

Table 6: Results for Test 0.

## 10.1   Test 0 - No Memory Stress

In this test, the container that receives HTTP requests from the load generator is given enough memory so that no memory stress situation occurs. This establishes upper-bound (best case) reference statistics.

The OpenVZ user beancounter barriers and limits for the container that will receive requests from the load generator are shown in Table 5a, which are the values for the 128-MiB profile described earlier. The format for the beancounter values is `barrier:limit` where 100M means 100 mebibytes ($100 \times 2^{20}$ bytes). The OpenVZ User Guide states that the value for unused barriers or limits should be set to the special value "unlimited" (an alias for the maximum representable unsigned integer

**Apache HTTP Server Settings**

| | |
|---|---|
| StartServers | 4 |
| MinSpareServers | 2 |
| MaxSpareServers | 4 |
| MaxClients | 128 |
| KeepAlive | On |
| KeepAliveTimeout | 5 |

**Initial OpenVZ Settings**

| | |
|---|---|
| oomguarpages | 60M:unlimited |
| vmguarpages | 64M:unlimited |
| privvmpages | 64M:66M |

**JMeter Settings**

| | |
|---|---|
| Number of Threads (users) | 9 |
| Ramp-Up Period (in seconds) | 2 |
| Loop Count | 5 |

**Golondrina Settings**

| | |
|---|---|
| [server/policy/state] | |
| overload-mem-default | {'threshold':0.80} |
| [client] | |
| frequency | 10 |

Table 7: OpenVZ, Apache HTTP Server, JMeter and Golondrina settings for Test 1.

on the current processor) for future compatibility.

The Apache HTTP Server settings for this test are shown in Table 5b. The main parameters of interest are the `StartServers`, `MinSpareServers` and `MaxSpareServers` directives. The `KeepAliveTimeout` parameter is measured in seconds.

Table 5c shows the setting for JMeter and Table 5d shows the settings for Golondrina. The client `frequency` value is the delay, in seconds, between each collection of resource usage statistics.

The results for four test runs are given in Table 6 as well as the average values for the four runs. The Avg, Min and Max columns list the average, minimum and maximum response times for HTTP requests of PHP web pages as measured by JMeter, in milliseconds. The Std Dev column lists the standard deviation of the response times. The Err % column lists the percentage of HTTP requests that were unsuccessful (that is, the web server did not return a success code, or a success code was returned but the page did not contain a closing HTML tag). The throughput column lists the number of requests fulfilled per second. The fail count column lists the number of memory allocation attempts by user-space processes that were denied by the OpenVZ kernel during the test.

## 10.2   Test 1 - Unresolved Memory Stress

The test establishes lower-bound (worst case) reference statistics. The memory allocated to the container that receives HTTP requests has been reduced to the 64-MiB profile and a memory stress situation occurs in the container after about six

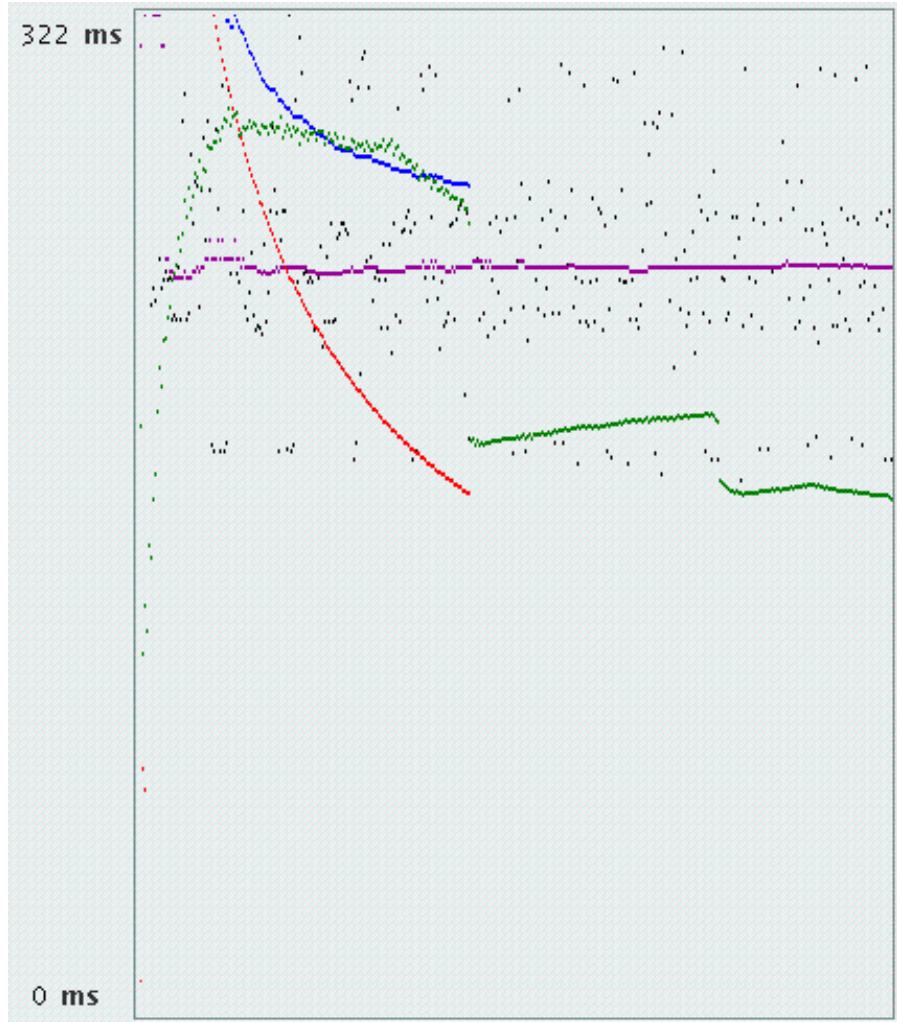

Figure 20: A Plot of the Test 1 results as a function of time. Time is represented along the horizontal axis while the response times for HTTP requests are shown along the vertical axis. The black dots represent the response times for individual HTTP requests while the blue dots represent the average response time, and the purple dots represent the median response time. The red dots represent the standard deviation while the green dots represent the throughput.

| Run | Avg | Min | Max | Std Dev | Err % | Throughput | Fail Count |
|-----|-----|-----|-----|---------|-------|------------|------------|
| a | 1402 | 167 | 94369 | 9173.26 | 0.28 | 2.8 | 748 |
| b | 1051 | 120 | 40411 | 4592.58 | 4.72 | 3.4 | 1136 |
| c | 996 | 43 | 38028 | 4302.32 | 3.89 | 4.1 | 1259 |
| d | 877 | 166 | 49494 | 5201.46 | 0.00 | 4.2 | 865 |
|  | 1082 | 124 | 55576 | 5817.40 | 2.22 | 3.6 | 1002 |

Table 8: Results for Test 1.

**Initial OpenVZ Settings**

```
oomguarpages    60M:unlimited
vmguarpages     64M:unlimited
privvmpages        64M:66M
```

**Apache HTTP Server Settings**

```
StartServers              4
MinSpareServers           2
MaxSpareServers           4
MaxClients              128
KeepAlive                On
KeepAliveTimeout          5
```

**JMeter Settings**

| Number of Threads (users) | 9 |
|---|---|
| Ramp-Up Period (in seconds) | 2 |
| Loop Count | 5 |

**Golondrina Settings**

```
[server/policy/state]
overload-mem-default    {'threshold':0.80}
[client]
frequency                            10
```

Table 9: OpenVZ, Apache HTTP Server, JMeter and Golondrina settings for Test 2.

JMeter test threads have started. The Golondrina memory management plug-in is disabled, leaving the memory stress situation unresolved. The settings for OpenVZ, the Apache HTTP Server, JMeter and Golondrina are given in Table 7. The results for this test are shown as a function of time in Figure 20 and summarized in Table 8.

## 10.3   Test 2 - Memory Stress Resolved Locally

In this test, the memory allocated to the container that receives HTTP requests has been reduced to the 64-MiB profile initially, but the Golondrina memory management plug-in has been enabled. The memory management plug-in applies the heuristic described in Section 4.1 and increases the memory allocated to the container to the 128-MiB profile if its normalized score exceeds the threshold value given in Table 9. The results for this test are summarized in Table 10.

| Run | Avg | Min | Max | Std Dev | Err % | Throughput | Fail Count |
|-----|-----|-----|-----|---------|-------|------------|------------|
| a | 833 | 47 | 34944 | 3992.46 | 3.89 | 4.3 | 1801 |
| b | 906 | 75 | 51716 | 4532.85 | 2.64 | 7.4 | |
| c | 790 | 168 | 51716 | 4955.03 | 0.00 | 4.1 | 330 |
| d | 933 | 169 | 58481 | 5515.03 | 0.00 | 3.8 | 297 |
| | 866 | 115 | 49214 | 4748.84 | 1.63 | 4.9 | 809 |

Table 10: Results for Test 2.

**Initial OpenVZ Settings**

```
oomguarpages   60M:unlimited
vmguarpages    64M:unlimited
privvmpages         64M:66M
```

**Apache HTTP Server Settings**

| | |
|---|---|
| `StartServers` | 4 |
| `MinSpareServers` | 2 |
| `MaxSpareServers` | 4 |
| `MaxClients` | 128 |
| `KeepAlive` | `On` |
| `KeepAliveTimeout` | 5 |

**JMeter Settings**

| | |
|---|---|
| Number of Threads (users) | 9 |
| Ramp-Up Period (in seconds) | 36 |
| Loop Count | 5 |

**Golondrina Settings**

```
[server/policy/state]
overload-mem-default   {'threshold':70}
[client]
frequency                            2
```

Table 11: OpenVZ, Apache HTTP Server, JMeter and Golondrina settings for Test 3.

## 10.4   Test 3 - Memory Stress Resolved with Migration

In this test, the memory allocated to the container that receives HTTP requests has been reduced to the 64-MiB profile and the Golondrina memory management plug-in has been enabled. Once the container receives several simultaneous HTTP requests, it becomes memory stressed. The memory management plug-in will try to increase the memory limits for the container on the current hardware node, but find that the hardware node does not have enough free memory to allow this action. Thus, the container is first migrated to another hardware node, and then it has its memory limits increased.

The JMeter ramp-up time for this test has been increased, and the Golondrina threshold and frequency values were decreased to allow Golondrina to react and migrate the container before it runs out of memory. As long as all the JMeter threads run at the same time at some point, then the same number of Apache processes should be created and thus the test should have the same memory footprint.

| Run | Avg | Min | Max | Std Dev | Err % | Throughput | Fail Count |
|---|---|---|---|---|---|---|---|
| a | 414 | 166 | 6526 | 643.89 | 0.28 | 5.3 | 283 |
| b | 376 | 165 | 3612 | 364.68 | 0.00 | 5.3 | 0 |
| c | 279 | 49 | 3516 | 315.07 | 36.67 | 5.9 | 2286 |
| d | 409 | 169 | 3760 | 354.03 | 0.00 | 5.4 | 0 |
|  | 370 | 137 | 4354 | 419.42 | 9.24 | 5.5 | 1143 |

Table 12: Results for Test 3.

## 10.5  Analysis

By comparing the minimum response times of HTTP requests for Tests 0-3, it can be seen that the minimum response time for runs without errors was about 166 milliseconds, whereas test runs with an error rate greater than 0.28% had a minimum response time of 75 milliseconds or less. The low minimum response times for test runs with higher error rates are likely due to error responses being generated quickly. The low minimum response times likely occurred early in test runs when only the first test thread was running–before the container became memory stressed. This likely contributed to the high standard deviation seen in the test runs in which the container experienced a memory stress.

The large response times that resulted when the container became memory stressed were due to the fact that the Apache HTTP Server parent process was unable to create additional child processes. In this case, new connections were queued and handled by existing child processes after their current connections were terminated.

In Test 1 there was a noticeable increase in the maximum response time and corresponding decrease in the throughput. Each run in this test also had a large fail count, meaning that memory allocation requests, such as calls to the C library's malloc or fork functions, were denied. Thus, the container was severely memory stressed. During each test the CPU usage on the hardware node hosting the stressed container was monitored to ensure it remained below 100 percent. Similarly, the network bandwidth used was monitored during each test and remained below 1.2 megabits per second. Thus, the decrease in performance of the website in this test was due to the memory shortage.

The four runs of Test 0 (no memory stress) and Test 1 (unresolved memory stress) produced an average throughput of 8.0 and 3.6 respectively. The four test runs of Test 2 (memory stress resolved locally) produced an average throughput of 4.9. Golondrina detected the memory stress, as expected, and increased the stressed container's memory limits to the 128-MiB profile. This resulted in a throughput value that was substantially better than the throughput for Test 1 (unresolved memory stress), but not as good as the throughput for Test 0 (no memory stress). In this test, an average of 314 memory allocation attempts were denied by the kernel. This value is again substantially better than 1002, the average fail count for runs in Test 1, but it indicates that the container was still severely stressed for a short period of time. Although Golondrina successfully detected and resolved the memory stress situation, Golondrina did not react quickly enough to prevent memory requests from being denied. The reaction time could be improved by increasing the JMeter ramp-up time, decreasing the delay in the gathering of resource usage statistics, or decreasing the threshold value for the memory management plug-in.

It was discovered during Test 3 that OpenVZ starts a process in the container as part of the checkpointing process for a live migration. This process counts towards the container's overall resource usage. If the container does not have enough memory to accommodate this process, the migration will fail. This suggests that all Golondrina resource management plug-ins need to take the cost of this process into consideration before attempting to migrate a container. Therefore, in this test the ramp-up time was increased, the threshold value for the memory management plug-in was decreased, and the delay between resource usage statistics gathering and analysis was also decreased, in order to minimize the likelihood of the migration failing due to a lack of memory in the container. If the migration failed for a test run, the run was repeated. A successful migration occurred during each of the four test runs shown.

There was a significant variance in the error rate between runs for Test 1, 2 and 3. One possible explanation for this is the synchronization, or lack thereof, between the moment when a resource stress occurred and the moment when it was first detected by Golondrina. These experiments show an interesting correlation between the HTTP request error rate and the memory allocation request fail count: as long as the fail count remained below approximately one thousand, there were nearly zero failed HTTP requests. Further investigation is needed to confirm this relationship.

## 11   Future Work

A detailed analysis of the performance enhancements that can be achieved by replicating a container that is memory stressed have yet to be completed.

The CPU resource management code from the first version of Golondrina has been included in the current version as a plug-in. It is expected that the ability of this plug-in to identify and dissipate CPU resource stress situations will be similar to the results presented in [7], but a detailed performance analysis has yet to be carried out.

The interaction between the memory and CPU plug-ins when a container is experiencing both a memory and CPU resource stress at the same time has not yet been studied.

The following are some deficiencies that have been identified in the current version of Golondrina, and may be fixed in future prototypes:

- There is currently no mechanism to reclaim resources when the workload for a container and its replicas, if any, decreases.

- Currently when migrations are performed, a container's whole private area is copied from one hardware node to another. Using remote storage to host containers' private areas could considerably reduce the amount of time consumed by migrations.

- All decisions regarding resource allocation to containers are currently performed by plug-ins in the Golondrina server component. The system may be able to respond more efficiently to resource stress situations, increasing overall performance, if an escalation procedure were defined. This way some resource allocation decisions could be performed at the hardware node level. This could be achieved without modifying the system architecture.

- The current memory management plug-in does not incorporate resource usage prediction. Resource usage patterns of production systems could be studied and a strategy for resource usage prediction could be developed.

## 12   Conclusion

This thesis presented a system called Golondrina that performs dynamic resource management among a cluster of hardware nodes running operating-system level virtualization software. Different models of virtualization were discussed, as well as other systems that have a purpose similar to Golondrina.

The resource allocation mechanisms of OpenVZ, the particular virtualization technology used by Golondrina, were discussed and the current heuristic used by the memory management plug-in to detect and dissipate memory stress situations was presented.

The architecture of Golondrina was discussed and the functionality of the system was validated using a series of experimental tests. In an analysis of the results, it was discovered that live migration incurs a substantial memory usage overhead that had not been considered. Nevertheless, these tests demonstrated that Golondrina is successful at performing the duties for which it was intended.

Finally, a brief summary of experiments that have yet to be performed was given along with some future enhancements that may be made to the system.